\documentclass{lmcs}
\pdfoutput=1

\usepackage{lastpage}
\lmcsdoi{17}{2}{10}
\lmcsheading{}{\pageref{LastPage}}{}{}%
{May~03,~2019}{Apr.~22,~2021}{}

\usepackage[utf8]{inputenc}

\usepackage{microtype}

\usepackage{tablefootnote}

\usepackage{amsmath}
\usepackage{amsthm}
\usepackage{amssymb}
\usepackage{xspace}

\usepackage{booktabs}
\usepackage{tikz}

\usetikzlibrary{matrix,arrows,automata}
\usetikzlibrary{shapes,positioning,calc,arrows,snakes}

\tikzstyle{vertex}=[draw,inner sep=0cm]
\tikzstyle{reach}=[regular polygon,regular polygon sides=4, minimum size=1.0cm, vertex]
\tikzstyle{safety}=[regular polygon,regular polygon sides=3, minimum size=0.8cm, vertex]
\tikzstyle{switch}=[circle, minimum size=0.8cm, vertex]
\tikzstyle{gadget}=[minimum width=2.0cm, minimum height=0.8cm, vertex]

\usetikzlibrary{matrix,arrows,shapes,positioning,calc,snakes,decorations.markings}

\usetikzlibrary{arrows.meta}
\tikzset{>={Latex[width=1.5mm,length=1.5mm]}}

\pgfdeclarelayer{background}
\pgfsetlayers{background,main}

\DeclareMathOperator{\ord}{Ord}
\DeclareMathOperator{\mru}{MRU}
\newcommand{\coun}{\ensuremath{C}}
\DeclareMathOperator{\upd}{Upd}
\DeclareMathOperator{\suc}{Succ}
\DeclareMathOperator{\bal}{Bal}
\DeclareMathOperator{\rep}{Rep}
\DeclareMathOperator{\out}{Out}
\DeclareMathOperator{\successors}{Succ}

\newcommand{\nats}{\ensuremath{\mathbb N}\xspace}
\newcommand{\vr}{\ensuremath{V_{\text{R}}\xspace}}
\newcommand{\vs}{\ensuremath{V_{\text{S}}\xspace}}
\newcommand{\vswi}{\ensuremath{V_{\text{Swi}}\xspace}}

\newcommand{\gadget}{\textsc{Gadget}\xspace}

\usepackage{cite}

\renewcommand{\L}{\ensuremath{\mathtt{L}}\xspace}
\newcommand{\NL}{\ensuremath{\mathtt{NL}}\xspace}
\newcommand{\PL}{\ensuremath{\mathtt{PL}}\xspace}

\renewcommand{\P}{\ensuremath{\mathtt{P}}\xspace}
\newcommand{\NP}{\ensuremath{\mathtt{NP}}\xspace}
\newcommand{\coNP}{\ensuremath{\mathtt{coNP}}\xspace}
\newcommand{\NPcapcoNP}{\ensuremath{\NP \cap \coNP}\xspace}
\newcommand{\UPcapcoUP}{\ensuremath{\UP \cap \coUP}\xspace}
\newcommand{\UP}{\ensuremath{\mathtt{UP}}\xspace}
\newcommand{\coUP}{\ensuremath{\mathtt{coUP}}\xspace}
\newcommand{\CLS}{\ensuremath{\mathtt{CLS}}\xspace}
\newcommand{\UEOPL}{\ensuremath{\mathtt{UEOPL}}\xspace}
\newcommand{\PLS}{\ensuremath{\mathtt{PLS}}\xspace}
\newcommand{\AP}{\ensuremath{\mathtt{AP}}\xspace}
\newcommand{\PPAD}{\ensuremath{\mathtt{PPAD}}\xspace}
\newcommand{\PSPACE}{\ensuremath{\mathtt{PSPACE}}\xspace}

\newcommand{\EXPTIME}{\ensuremath{\mathtt{EXPTIME}}\xspace}
\newcommand{\APSPACE}{\ensuremath{\mathtt{APSPACE}}\xspace}

\newcommand{\AUCSPACE}{\ensuremath{\mathtt{AUC\text{-}SPACE}(\log(n))}\xspace}

\title{Reachability Switching Games}

\author[J.~Fearnley]{John Fearnley\rsuper{a}}	

\author[M.~Gairing]{Martin Gairing\rsuper{a}}	
\address{\lsuper{a}University of Liverpool, Liverpool, United Kingdom}	
\email{john.fearnley@liverpool.ac.uk, gairing@liverpool.ac.uk, rahul.savani@liverpool.ac.uk}  

\author[M.~Mnich]{Matthias Mnich\rsuper{b}}

\author[R.~Savani]{Rahul Savani\rsuper{a}}	

\address{\lsuper{b}Hamburg University of Technology, Institute for Algorithms and Complexity, Hamburg, Germany}
\email{matthias.mnich@tuhh.de}

\subjclass{Theory of computation $\rightarrow$  Logic $\rightarrow$ Logic and verification}
\titlecomment{A preliminary version of this paper appeared at ICALP 2018~\cite{FGMS18b}.}

\begin{document}

\maketitle

\begin{abstract}
  We study the problem of deciding the winner of reachability switching games for 
  zero-, one-, and two-player variants.
  Switching games provide a deterministic analogue of stochastic games.
  We show that the zero-player case is \NL-hard, the one-player case is \NP-complete, and that the two-player case is \PSPACE-hard and in \EXPTIME.
  For the zero-player case, we also show \P-hardness for a succinctly-represented model that maintains the upper bound of $\NP \cap \coNP$.
  For the one- and two-player cases, our results hold in both the natural, explicit model and succinctly-represented model.
  Our results show that the switching variant of a game is harder in complexity-theoretic terms
  than the corresponding stochastic version.
\end{abstract}

\section{Introduction}
\label{sec:introduction}

A \emph{switching system} (also known as a Propp machine) attempts to replicate
the properties of a random system in a deterministic way~\cite{HP10}. It does so
by replacing the nodes of a Markov chain with \emph{switching nodes}. Each
switching node maintains a queue over its outgoing edges. When the system
arrives at the node, it is sent along the first edge in this queue, and that
edge is then sent to the back of the queue. In this way, the switching node
ensures that, after a large number of visits, each outgoing edge
is used a roughly equal number of times.

The Propp machine literature has focussed on \emph{many-token} switching systems
and has addressed questions such as how well these systems emulate Markov
chains. Recently, Dohrau et al.~\cite{DGKMW17} initiated the study of
\emph{single-token} switching systems and found that the reachability problem
raised interesting complexity-theoretic questions. Inspired by that work, we
study the question \emph{how hard is it to model check single-token switching
systems?} 
A switching node is a simple example of a fair scheduler, and thus it is 
natural to consider model checking of switching systems.
We already have a good knowledge about the complexity of model
checking Markovian systems, but how does this change when we instead use
switching nodes?

\paragraph{Our contribution.}
All work so far has studied zero-player switching systems. In this
paper, we initiate the study of model checking in switching systems, which
naturally leads to one- and two-player switching systems. We focus on one-token
\emph{reachability} problems, one of the simplest model checking tasks for a
switching system. This corresponds to determining the winner of a
\emph{two-player reachability switching game}. We study zero-, one-, and
two-player variants of these games, which correspond to switching versions of
Markov chains, Markov decision processes~\cite{Put94}, and simple stochastic
games~\cite{Con92}, respectively. Our results are summarised in
\Cref{tab:summary}.

The main message of the paper is that deciding reachability in one- and
two-player switching games is harder than deciding reachability in
Markovian systems. Specifically, we show that deciding the winner of a
one-player game is \NP-complete, and that the problem of
deciding the winner of a two-player game is \PSPACE-hard and in
\EXPTIME. 

We also study the complexity of zero-player games, where we show hardness
results. For the standard model of switching systems, which we call
\emph{explicit games}, we are able to show a lower bound of \NL-hardness, which
is still quite far from the known upper bound of \UPcapcoUP \cite{GHHKMS18}. We
also show that if one extends the model by allowing the switching order to be
represented in a concise way, then a stronger lower bound of \P-hardness can be
shown. We call these concisely represented games \emph{succinct games}, and we
are also able to show upper bounds for succinct zero-player games that match the
known upper bounds for explicit zero-player games. Furthermore, all of our other
results for one and two-player games, both upper and lower bounds, still apply
to succinct games.


\begin{table}[h] 
\noindent\begin{center}\scalebox{0.75}{%
\renewcommand{\tabcolsep}{7pt}%
\renewcommand{\arraystretch}{1.3}%
  \begin{tabular}{l@{\hspace{4pt}}lll}
      \toprule
	  & \textbf{Markovian} & \textbf{Switching (explicit)} & \textbf{Switching (succinct)} \\

      \midrule
	  0-pl. & \PL-compl.\tablefootnote{\PL, or probabilistic \L, is the
	  class of languages recognizable by a polynomial-time logarithmic-space
      randomized machine with probability $>$ 1/2; there is a straight-forward
      polynomial-time inter-reduction between this decision problem for polynomial-time logarithmic-space
      randomized machines and Markov chains.} &
	  \NL-hard \hfill (Thm.~\ref{thm:NLhard}) 
	  & \P-hard \hfill (Thm.~\ref{thm:phard})  \\ 
	  && in \UEOPL, \CLS, and \UPcapcoUP \hfill \cite{GHHKMS18}
	  & in \UEOPL and \CLS \hfill (Thm.~\ref{thm:ueopl})\\
      \midrule
	  1-pl. & \P-compl. \hfill \cite{PT87} 
                           & \NP-compl. \hfill (Thm.~\ref{thm:np} and~\ref{thm:nphard})
                           & \NP-compl. \hfill (Thm.~\ref{thm:np} and~\ref{thm:nphard})\\

      \midrule
	  2-pl. & \AUCSPACE-compl.\tablefootnote{\AUCSPACE is the class of languages
	  			accepted by log-space bounded randomized alternating Turing machines.
	  		    It was defined by Condon~\cite{Con89}.}
	  			\hfill \cite{Con89}
			   & \PSPACE-hard \hfill (Thm.~\ref{thm:2p}) 
			   & \PSPACE-hard \hfill (Thm.~\ref{thm:2p}) \\
	  		   & in \NPcapcoNP \hfill \cite{Con92}
	  		   & in \EXPTIME \hfill (Thm.~\ref{thm:2player-upper}) 
	  		   & in \EXPTIME \hfill (Thm.~\ref{thm:2player-upper})\\
      \bottomrule
    \end{tabular}
}
\end{center}
\caption{A summary of our results.}
\label{tab:summary}
\end{table}

\noindent 
For the explicit zero-player case
the first bound was an \NPcapcoNP upper bound given by 
Dohrau et al.~\cite{DGKMW17}, and a \PLS upper bound was then given by
Karthik~\cite{Kar17}.
The \UEOPL, \CLS, and \UPcapcoUP upper bounds, which subsume the two earlier bounds, were
given
by G\"{a}rtner et al.~\cite{GHHKMS18}, who also produced a $O(1.4143^n)$-time
algorithm for solving explicit zero-player games. All the other upper and lower
bounds in the table are proved in this paper. 
 
Finally, we address the memory requirements of winning strategies in
reachability switching games. It is easy to see that winning strategies exist
that use exponential memory. These strategies simply remember the current switch
configuration of the switching nodes, and their existence can be proved by
blowing up a switching game into an exponentially sized reachability game, and
then following the positional winning strategies from that reachability game. This
raises the question of whether there are winning strategies that use less than
exponential memory. We answer this negatively, by showing that the reachability
player may need $2^{\Omega(n)}$ memory states to win a one-player reachability
switching game, and that both players may need to use $2^{n/6}$ memory
states to win a two-player game. 

\paragraph{Related work.}
%
While we are the first to study switching games with multiple players,
zero-player switching systems (so far mainly with multiple tokens) are part of a
research thread at the intersection of computer science and physics. These 
zero-player switching systems are also known as \emph{deterministic
	random walks}, \emph{rotor-router walks}, the~\emph{Eulerian walkers
model}~\cite{PDDK96} and \emph{Propp
machines}~\cite{HLMPP08,CDST07,CDFS10,DF09,CS06,HP10}.  
These systems have been studied in the context of
derandomizing algorithms and pseudorandom simulation, and in particular have
received a lot of attention in the context of load balancing~\cite{FGS12,AB13}.
Most work on switching systems has focused on how well multi-token
switching systems simulate Markov chains.

The idea of studying
\emph{single-token} reachability is due to Dohrau et al.~\cite{DGKMW17}, who
introduced the problem under the name {\sc Arrival}. Subsequent work has shown that
{\sc Arrival} lies in the complexity classes \PLS~\cite{Kar17}, \CLS~\cite{GHHKMS18,FGHS21},
and \UEOPL~\cite{FGMS20}. Recently, a deterministic sub-exponential algorithm has
been given for solving the {\sc Arrival} problem~\cite{GHH21}.

We study model checking for single-token switching systems. There is extensive
literature on model checking stochastic systems, known as \emph{probabilistic
model checking}, which is an central topic in the field of formal verification.
Markov decision processes~\cite{Put94} and simple stochastic games~\cite{Con92}
are important objects of study in probabilistic model checking. Probabilistic
model checking is now a mature topic, with tools like PRISM~\cite{KNP11}
providing an accessible interface to the research that has taken place.

Other notions of switching nodes have been studied.
Katz et al.~\cite{KRW12}, Groote and Ploeger~\cite{GP09}, and
others~\cite{GP09,Mei89,Rei09}, considered \emph{switching
graphs}; these are graphs in which certain vertices (switches) have exactly one
of their two outgoing edges activated.  However, the activation of the alternate
edge does not occur when a vertex is traversed by a run; this is the key
difference to switching games in this paper.


\paragraph{Outline of the paper.}
The remainder of the paper is structured as follows. In \Cref{sec:prelim},
we formally introduce reachability switching games (RSGs).
In \Cref{sec:oneplayer}, we deal with one-player RSGs and show that 
solving them is \NP-complete. We end this section by analysing the 
memory requirements of winning strategies in one-player RSGs.
In \Cref{sec:twoplayer}, we deal with two-player RSGs, first showing
that they can be solved in \EXPTIME and then showing a \PSPACE lower bound.
We end this section by analysing the 
memory requirements of winning strategies in two-player RSGs.
In \Cref{sec:zeroplayer}, we deal with zero-player RSGs,
first showing that solving them is \NL-hard for the explicit representation
and \P-hard for the succinct representation, and then showing an upper bound
of \UEOPL. Finally, in \Cref{sec:fw}, we discuss open problems and further 
work.

\section{Preliminaries}
\label{sec:prelim}

In this section, we define \emph{two-player} reachability switching games between
a reachability player and a safety player.
One-player games refer to games in which the safety player has no vertices, and
zero-player games refer to games in which both the reachability and safety
players have no vertices.
The zero-player case corresponds to the problem studied by 
Dohrau et. al.~\cite{DGKMW17}, which they call {\sc Arrival}.

A reachability switching game (RSG) is defined by a tuple $(V, E, \vr, \vs, \vswi,
\ord, s, t)$, where $(V, E)$ is a finite directed graph, and \vr, \vs, $\vswi$ partition
$V$ into \emph{reachability vertices}, \emph{safety vertices}, and
\emph{switching vertices}, respectively. The reachability vertices $\vr$ are
controlled by the reachability player, the safety vertices $\vs$ are
controlled by the safety player, and the switching vertices~$\vswi$ are
not controlled by either player, but instead follow a predefined \emph{switching
order}. The function $\ord$ defines this switching order: for each
switching vertex $v \in \vswi$, we have that $\ord(v) = \langle u_1, u_2, \dots,
u_k \rangle$ where we have that $(v, u_i) \in E$ for all $u_i$ in the sequence.
Note that a particular vertex $u$ may appear more than once in the sequence.
The vertices $s, t \in V$ specify
\emph{source} and \emph{target} vertices for the game. 

A \emph{state} of the game is defined by a tuple $(v, \coun)$, where $v$ is a
vertex in $V$, and $\coun: \vswi \rightarrow \nats$ is a function that assigns
a number to each switching vertex, which represents how far that vertex has
progressed through its switching order. Hence, it is required that $\coun(u)
\le |\ord(v)| - 1$, since the counts specify an index to the sequence $\ord(v)$.

When the game is at a state $(v, \coun)$ with $v \in \vr$ or $v \in \vs$, then
the respective player chooses an outgoing edge at $v$, and the count function
does not change.
For states $(v, \coun)$ with $v \in \vswi$, the successor state is determined by
the count function. More specifically, we define $\upd(v, \coun): \vswi \rightarrow
\nats$ so that for each $u \in \vswi$ we have
$\upd(v, \coun)(u) = \coun(u)$ if $v\not= u$, and
$\upd(v, \coun)(u) = 
(\coun(u) + 1) \bmod |\ord(u)|$ otherwise.
This function increases the count at~$v$ by $1$, and wraps around to $0$ if the
number is larger than the length of the switching order at~$v$. Then, the successor
state of $(v, \coun)$, denoted as $\suc(v, \coun)$ is $(u, \upd(v, \coun))$,
where $u$ is the element at position $C(v)$ in $\ord(v)$.

A \emph{play} of the game is a (potentially infinite) sequence of states $(v_1, C_1), (v_2, C_2), \dots$
with the following properties:
\begin{enumerate}
  \item $v_1 = s$ and $C_1(v) = 0$ for all $v \in \vswi$;
  \item If $v_i \in \vr$ or $v_i \in \vs$ then $(v_i, v_{i+1}) \in E$ and $C_i = C_{i+1}$;
  \item If $v_i \in \vswi$ then $(v_{i+1}, C_{i+1}) = \suc(v_i, C_i)$;
  \item If the play is finite, then the final state $(v_n, C_n)$ must either satisfy $v_n = t$, or $v_n$ must have no outgoing edges.
\end{enumerate}
A play is \emph{winning for the reachability player} if it is finite and the final state is 
the target vertex. 
For a zero-player game there is a single play, which is called a ``run profile''
by Dohrau et al.~\cite{DGKMW17}.

A (deterministic, history dependent) \emph{strategy for the reachability player}
is a function that maps each play prefix $(v_1, C_1), (v_2, C_2),\hdots,(v_k,
C_k)$, with $v_k \in \vr$, to an outgoing edge of~$v_k$.
A play $(v_1, C_1), (v_2, C_2),
\dots$ is \emph{consistent} with a strategy if, whenever $v_i \in \vr$, we have
that $(v_i,v_{i+1})$ is the edge chosen by the strategy.
Strategies for the safety player are defined analogously. 
A strategy is \emph{winning} if all plays consistent
with it are
winning. 

\paragraph{\bf The representation of the switching order.} 
Recall that $\ord(v) = \langle u_1, u_2, \dots, u_k \rangle$ gives a sequence of
outgoing edges for every switching vertex. We consider two possible ways of
representing $\ord(v)$ in this paper. In \emph{explicit} RSGs, $\ord(v)$ is
represented by simply writing down the sequence $\langle u_1, u_2, \dots, u_k
\rangle$. 

Natural switching orders may contain runs of identical next nodes, e.g.,
$\ord(v)$ may contain $u_\ell, u_{\ell+1}, \dots, u_{\ell+m}$ with $u_\ell = u_{\ell+1} = \cdots = u_{\ell+m}$.
In that case, if $m$ is large the explicit representation is wasteful.
Other, less wasteful representations are possible which give games
with a different computational complexity to explicitly represented games.
Motivated by this, we also consider games in which $\ord(v)$ is written down in
a more concise way, which we call \emph{succinct} RSGs.
In succinct games, for each switching vertex $v$, we have a sequence of pairs
$\langle (u_1, t_1), (u_2, t_2), \dots, (u_k, t_k) \rangle$, where each $u_i$ is
a vertex with  $(v, u_i) \in E$, and each $t_i$ is a natural number. The idea is
that $\ord(v)$ should contain $t_1$ copies of $u_1$, followed by $t_2$ copies of
$u_2$, and so on. So, if $\rep(u, t)$ gives the sequence containing $t$ copies
of $u$, and if $\cdot$ represents sequence concatenation, then 
\begin{equation*}
\ord(v) = \rep(u_1, t_1) \cdot \rep(u_2, t_2) \cdot \; \dots \; \cdot \rep(u_k,
t_k).
\end{equation*}
Any explicit game can be
written down in the succinct encoding by setting all $t_i = 1$. Note, however,
that in a succinct game $\ord(v)$ may have exponentially many elements, even if
the input size is polynomial, since
each $t_i$ is represented in binary.

\section{One-player reachability switching games}
\label{sec:oneplayer}
In this section we consider one-player RSGs, i.e., where $\vs = \emptyset$.

\subsection{Containment in \NP}

We show that deciding whether the reachability player wins a one-player
RSG is in \NP. Our proof holds for both explicit and
succinct games.  The proof uses \emph{controlled switching flows}. These extend
the idea of switching flows, which were used by Dohrau et al.~\cite{DGKMW17} to show
containment of the zero-player reachability problem in \NPcapcoNP.

\paragraph{\bf Overview.}

A controlled switching flow assigns an integer to every edge of the
game. 
The flow is required to satisfy \emph{balance} constraints
that ensure that the flow entering a vertex is equal to the flow leaving that
vertex (except for the source and target nodes).
It is also required to satisfy a \emph{switching} constraint, which will ensure
that switching nodes send the correct amount of flow 
(i.e., consistent with the switching order) 
to each of their successors.

\tikzstyle{triangle}=[regular polygon, regular polygon sides=3]
\tikzstyle{vertex}=[draw, thick, inner sep=0cm]
\tikzstyle{max}=[minimum size=0.8cm, vertex]
\tikzstyle{min}=[minimum size=1.1cm, vertex, triangle]
\tikzstyle{avg}=[minimum size=0.8cm, vertex, circle]
\tikzstyle{token}=[minimum size=0.4cm, vertex, circle, fill=token, draw=token]
\tikzstyle{maxtarget}=[minimum size=0.8cm, vertex, fill=maxwin]
\tikzstyle{mintarget}=[minimum size=1.1cm, vertex, triangle, fill=minwin]
 
\tikzstyle{normal}=[->, thick]
\tikzstyle{curved}=[normal, bend left=-30]

\begin{figure}[ht]
\begin{minipage}[h]{0.32\textwidth}
\begin{center}
\resizebox{1\linewidth}{!}{
\begin{tikzpicture}

\node[max] (source) {$s$};
\node[avg, right of=source, node distance=2cm] (avg1) {};
\node[max, right of=avg1, node distance=3cm] (max2) {$u$};
\node[avg, below of=max2, node distance=3cm] (avg2) {};
\node[max, below of=avg1, node distance=3cm] (deadend) {};
\node[max, right of=avg2, node distance=2cm] (sink) {$t$};


\path (source) edge [normal] (avg1);

\path (avg1) edge [curved,normal] node[at start,xshift=2mm,yshift=1mm] {$1$}  (max2);
\path (avg1) edge [curved,normal] node[at start, right,yshift=-1mm] {$2$} (avg2);
\path (avg1) edge [normal] node[at start,left,yshift=-1mm] {$3$} (deadend);

\path (max2) edge [curved, normal] (avg2);
\path (max2) edge [curved, normal] (avg1);

\path (avg2) edge [curved,normal] node[at start,left,yshift=2.5mm,xshift=2mm] {$1$}  (max2);
\path (avg2) edge [normal] node[at start,below,xshift=1mm] {$2$} (sink);


\path (source) edge [normal] node[above,draw,thin,yshift=1mm] {$1$} (avg1);

\path (avg1) edge [curved,normal] node[below,draw,thin,yshift=-1mm] {$1$}  (max2);
\path (avg1) edge [curved,normal] node[right,draw,thin,yshift=2.5mm] {$1$} (avg2);
\path (avg1) edge [normal] node[left,draw,thin,xshift=-1mm] {$0$} (deadend);

\path (max2) edge [curved, normal] node[left,draw,thin,xshift=-1mm] {$1$} (avg2);
\path (max2) edge [curved, normal] node[below,draw,thin,yshift=-1mm] {$1$} (avg1);

\path (avg2) edge [curved,normal] node[right,draw,thin,xshift=1mm] {$1$}  (max2);
\path (avg2) edge [normal] node[above,draw,thin,yshift=1mm] {$1$} (sink);

\end{tikzpicture}
}
\end{center}
\end{minipage}
\hfill
\begin{minipage}[h]{0.32\textwidth}
\begin{center}
\resizebox{1\linewidth}{!}{
\begin{tikzpicture}

\node[max] (source) {$s$};
\node[avg, right of=source, node distance=2cm] (avg1) {};
\node[max, right of=avg1, node distance=3cm] (max2) {$u$};
\node[avg, below of=max2, node distance=3cm] (avg2) {};
\node[max, below of=avg1, node distance=3cm] (deadend) {};
\node[max, right of=avg2, node distance=2cm] (sink) {$t$};


\path (source) edge [normal] (avg1);

\path (avg1) edge [curved,normal] node[at start,xshift=2mm,yshift=1mm] {$1$}  (max2);
\path (avg1) edge [curved,normal] node[at start, right,yshift=-1mm] {$2$} (avg2);
\path (avg1) edge [normal] node[at start,left,yshift=-1mm] {$3$} (deadend);

\path (max2) edge [curved, normal] (avg2);
\path (max2) edge [curved, normal] (avg1);

\path (avg2) edge [curved,normal] node[at start,left,yshift=2.5mm,xshift=2mm] {$1$}  (max2);
\path (avg2) edge [normal] node[at start,below,xshift=1mm] {$2$} (sink);


\path (source) edge [normal] node[above,draw,thin,yshift=1mm] {$1$} (avg1);

\path (avg1) edge [curved,normal] node[below,draw,thin,yshift=-1mm] {$1$}  (max2);
\path (avg1) edge [curved,normal] node[right,draw,thin,yshift=2.5mm] {$0$} (avg2);
\path (avg1) edge [normal] node[left,draw,thin,xshift=-1mm] {$0$} (deadend);

\path (max2) edge [curved, normal] node[left,draw,thin,xshift=-1mm] {$2$} (avg2);
\path (max2) edge [curved, normal] node[below,draw,thin,yshift=-1mm] {$0$} (avg1);

\path (avg2) edge [curved,normal] node[right,draw,thin,xshift=1mm] {$1$}  (max2);
\path (avg2) edge [normal] node[above,draw,thin,yshift=1mm] {$1$} (sink);

\end{tikzpicture}
}
\end{center}
\end{minipage}
\hfill
\begin{minipage}[h]{0.32\textwidth}
\begin{center}
\resizebox{1\linewidth}{!}{
\begin{tikzpicture}

\node[max] (source) {$s$};
\node[avg, right of=source, node distance=2cm] (avg1) {};
\node[max, right of=avg1, node distance=3cm] (max2) {$u$};
\node[avg, below of=max2, node distance=3cm] (avg2) {};
\node[max, below of=avg1, node distance=3cm] (deadend) {};
\node[max, right of=avg2, node distance=2cm] (sink) {$t$};


\path (source) edge [normal] (avg1);

\path (avg1) edge [curved,normal] node[at start,xshift=2mm,yshift=1mm] {$1$}  (max2);
\path (avg1) edge [curved,normal] node[at start, right,yshift=-1mm] {$2$} (avg2);
\path (avg1) edge [normal] node[at start,left,yshift=-1mm] {$3$} (deadend);

\path (max2) edge [curved, normal] (avg2);
\path (max2) edge [curved, normal] (avg1);

\path (avg2) edge [curved,normal] node[at start,left,yshift=2.5mm,xshift=2mm] {$1$}  (max2);
\path (avg2) edge [normal] node[at start,below,xshift=1mm] {$2$} (sink);


\path (source) edge [normal] node[above,draw,thin,yshift=1mm] {$1$} (avg1);

\path (avg1) edge [curved,normal] node[below,draw,thin,yshift=-1mm] {$1$}  (max2);
\path (avg1) edge [curved,normal] node[right,draw,thin,yshift=2.5mm] {$0$} (avg2);
\path (avg1) edge [normal] node[left,draw,thin,xshift=-1mm] {$0$} (deadend);

\path (max2) edge [curved, normal] node[left,draw,thin,xshift=-1mm] {$3$} (avg2);
\path (max2) edge [curved, normal] node[below,draw,thin,yshift=-1mm] {$0$} (avg1);

\path (avg2) edge [curved,normal] node[right,draw,thin,xshift=1mm] {$2$}  (max2);
\path (avg2) edge [normal] node[above,draw,thin,yshift=1mm] {$1$} (sink);

\end{tikzpicture}
}
\end{center}
\end{minipage}
\caption{Three examples of controlled switching flows for the same game. The flows
are given inside squares on the edges; the other numbers at the start of switching
edges indicate the order of these switching edges. The left and centre examples correspond
to plays, the rightmost example does not.}
\label{fig:csf}
\end{figure}

Intuitively, one might think of a switching flow as counting the number of times
that an edge is used during a play. Indeed, if we take a winning play for the
reachability player, and construct a flow by counting the number of times that
each edge is used, then we will obtain a switching flow. The leftmost two
diagrams in \Cref{fig:csf} show two possible controlled switching flows
for a game. In our diagramming notation, vertices controlled by the reachability
player are represented as squares, while switching nodes are represented as
circles. The switching order is shown by the numbers attached to the outgoing
edges of the switching nodes. 

The left hand diagram shows a controlled switching flow that corresponds to the
player using the downwards edge from $u$ once, and the leftwards edge from $u$
once. The middle diagram shows a controlled switching flow that corresponds to
the player using the downwards edge twice. Note that both strategies will result
in the play reaching $t$.

Note that in the left hand diagram the order in which the edges are used is
irrelevant: the player will win no matter which edge is used first. We will show
that this is always the case. Given a controlled switching flow, the player can
adopt a \emph{marginal} strategy that simply uses each edge the proscribed
number of times, and this is sufficient to ensure that the player will win.

One interesting feature of switching flows, pointed out by 
Dohrau et al.~\cite{DGKMW17}, is that there can exist \emph{false} switching
flows. These flows \emph{do not} correspond to actual plays of the game. The
rightmost diagram in \Cref{fig:csf} shows a false controlled switching
flow. Note that no possible play of the game uses the downwards edge three times
while using the leftwards edge zero times, and so there is one unit of ``false
flow'' between $u$ and the vertex below it. 

This does not affect our results in this section, since if the player follows a
marginal strategy they will still win, though they will not be able to use the
edge three times.
However, we will have to deal
with false flows when we show UEOPL containment for zero-player succinct games
in \Cref{sec:ueoplzp}. 



\paragraph{\bf Controlled switching flows.}

Formally, a \emph{flow} is a function $F: E \rightarrow \nats$ that assigns a
natural number to each edge in the game.
For each vertex $v$, we define 
  $\bal(F, v) = \sum_{(v, u) \in E} F(v, u) - \sum_{(w, v) \in E} F(w, v)$,
which is the difference between the outgoing and incoming flow at $v$. For each
switching node $v \in \vswi$, let $\successors(v)$ denote the set of vertices
that appear in $\ord(v)$, and for
each index $i \le |\ord(v)|$ and each vertex $u \in \successors(v)$,
let $\out(v, i, u)$ be the number of times that $u$ appears in the first $i$
entries of $\ord(v)$. In other words, $\out(v, i, u)$ gives the amount of flow
that should be sent to $u$ if we send exactly $i$ units of flow into $v$.

A flow $F$ is a \emph{controlled switching flow} if it satisfies the following constraints:
\begin{enumerate}
  \item The source vertex $s$ satisfies $\bal(F, s) = 1$, and the target vertex $t$
satisfies $\bal(F, t) = -1$.
  \item Every vertex $v$ other than $s$ or $t$ satisfies $\bal(F, v) = 0$.
  \item Let $v \in \vswi$ be a switching node, $k = |\ord(v)|$, and let $I =
\sum_{(u, v) \in E} F(u, v)$ be the total amount of flow incoming to $v$. Define
$p$ to be the largest integer such that $p \cdot k \le I$ (which may be $0$),
and $q = I \bmod k$. For every vertex $w \in \successors(v)$ we have that $F(v,
w) = p \cdot \out(v, k, w) + \out(v, q, w)$. 

\end{enumerate}
The first two constraints ensure that $F$ is a flow from $s$ to $t$,
while the final constraint ensures that the flow respects the switching order at
each switching node. 
Note that there are no constraints on how the flow is split
at the nodes in $\vr$.
For each flow $F$, we define the size of~$F$ to be $\sum_{e \in E} F(e)$. A flow of size $k$ can be written down using at most $|E| \cdot \log k$ bits.

\paragraph{\bf Marginal strategies.}
A \emph{marginal} strategy for the reachability player is defined by a function $M: E \rightarrow \nats$, which
assigns a target number to each outgoing edge of the vertices in~$\vr$.
The strategy ensures that each edge $e$ is used no more than $M(e)$ times.
That is, when the play arrives at a vertex $v \in \vr$, the strategy checks how
many times each outgoing edge of $v$ has been used so far, and selects an
arbitrary outgoing edge $e$ that has been used strictly less than $M(e)$ times.
If there is no such edge, then the strategy is undefined.

Observe that a controlled switching flow defines a marginal strategy for the reachability player.
We prove that this strategy always reaches the target.
\begin{lem}
\label{lem:mar2sw}
  If a one-player RSG has a controlled switching flow $F$, then any corresponding marginal strategy is winning for the reachability player.
\end{lem}

\begin{proof}
  The proof will be by induction on the size of $F$. 
  The base case is when $\sum_{e \in E} F(e) = 1$. The requirements of a controlled
switching flow imply that $F(s, t) = 1$, and all other edges have no flow at
all. If $s \in \vr$, then the corresponding marginal strategy is required to
choose the edge~$(s, t)$, and thus it is a winning strategy. If $s \in \vswi$,
then the balance requirement of a controlled switching flow ensures that $t$ is
the first vertex in $\ord(s)$, so the switching node will move to $t$, and the
reachability player will win the game.

There are two cases to consider for the inductive step. First, assume that 
$\sum_{e \in E} F(e) = i$, and that $s \in \vr$. Let $(s, v)$ be the outgoing
edge chosen by the marginal strategy (this can be any node that satisfies $F(s,
v) > 0$). If $G$ denotes the current game, then we can create a new switching
game~$G'$, which is identical to $G$, but where $v$ is the designated starting
node. Moreover, we can create a controlled switching flow $F'$ for $G'$ by
setting $F'(s, v) = F(s, v) - 1$ and leaving all other flow values unchanged.
Observe that all properties of a controlled switching flow continue to hold for
$F'$.
Since $\sum_{e \in E} F'(e) = i - 1$, the inductive hypothesis implies that the
marginal strategy that corresponds to $F'$ (which is consistent with the
marginal strategy for $F$) is winning for the reachability player.

The second case for the inductive step is when $\sum_{e \in E} F(e) = i$ and $s
\in \vswi$. Let $(s, v)$ be the first edge in $\ord(s)$, which is the edge that
the switching node will use. Again we can define a new game~$G'$ where the
starting node is $v$, and in which $\ord(s)$ has been rotated so that $v$
appears at the end of the sequence. We can define a controlled switching flow
$F'$ for~$G'$ where 
$F'(s, v) = F(s, v) - 1$ and all other flow values are unchanged. Observe that
$F'$ satisfies all conditions of a controlled switching flow, and in particular
that rotating $\ord(s)$ allows $s$ to continue to satisfy the balance constraint
on its outgoing edges. 
Again, since $\sum_{e \in E} F'(e) = i - 1$, the marginal strategy corresponding
to $F'$ (which is identical to the marginal strategy for $F$) is winning for the
reachability player. 
\end{proof}

In the other direction, if the reachability player has a winning strategy, then
we can prove that there exists a controlled switching flow, and moreover we can
give an upper bound on its size.

\begin{lem}
\label{lem:sw2mar}
If the reachability player has a winning strategy for a one-player RSG, then
that game has a controlled switching flow $F$, and the size of $F$ is at most $n
\cdot \ell^n$, where $n$ is the number of nodes in the game and $\ell = \max_{v \in
\vswi} |\ord(v)|$.
\end{lem}

\begin{proof}
Let $v_1, v_2, \dots, v_k$ be the play that is produced when the reachability
player uses his winning strategy. Define a flow $F$ so that $F(e)$ is the number
of times $e$ is used by the play.  We claim that~$F$ is a controlled switching
flow. In particular, since the play is a path through the graph starting at $s$
and ending at $t$, we will have $\bal(F, s) = 1$ and $\bal(F, t) = -1$, and we
will have $\bal(F, v) = 0$ for every vertex $v$ other than $s$ and $t$.
Moreover, it is not difficult to verify that the balance constraint will be
satisfied for every vertex $v \in \vswi$.

We now prove a bound on the size of the flow. 
First, observe that if a state $(v, \coun)$ appears twice in the play, then we
can modify the strategy to eliminate this cycle. Since the strategy is winning,
we know that the original play must be finite, and so we can apply the previous argument finitely many
times to produce a winning strategy that visits each state at most once.
Since the size of $F$ is equal to the number of steps in the play, we can upper
bound the size of $F$ by the number of distinct states. Recall that $C$ consists
of $|\vswi|$ numbers, and that $C(v)$ can take at most $|\ord(v)|$ different
values. So the number of possible
values for $C$ is at most $\ell^{|\vswi|}$, and so the number of possible states is
at most $|V| \cdot \ell^{|\vswi|} \le n \cdot \ell^n$.
\end{proof}

Combing the two previous lemmas yields the following corollary.

\begin{cor}
If the reachability player has a winning strategy for a one-player RSG, then he
also has a marginal winning strategy.
\end{cor}

Finally, we can show that solving a one-player RSG is in
\NP.

\begin{thm}
\label{thm:np}
Deciding the winner of an explicit or succinct one-player RSG is in \NP.
\end{thm}

\begin{proof}
By \Cref{lem:mar2sw} and~\Cref{lem:sw2mar}, the reachability player can win
if and only if the game has a
controlled switching flow of size at most $n \cdot \ell^n$. So, we can
non-deterministically guess a flow of size $n \cdot \ell^n$ and then
verify that it satisfies the requirements of a controlled switching flow. For
explicit games (where $\ell \in O(n)$) this can clearly be done in polynomial time.

For succinct games, first observe that, if $N$ denotes the input size of the
game, then $\ell \le 2^N$. Thus, the size of the flow is at most $n \cdot
2 ^ {N \cdot n}$. Since the flow is represented by a set of numbers, each of
which is written in binary, it can be
represented by at most $\log(n \cdot 2 ^ {N \cdot n})$ bits, which is polynomial in the
input size. 

Secondly, we note that the requirements of a switching flow can still be checked
in polynomial time, even for a succinct RSG. The first two requirements
give balance constraints that do not refer to the switching order, and thus can be
checked in polynomial time no matter whether the game is explicit or succinct.

The third requirement of a controlled switching flow does refer to the switching
order. It requires us to check, for each switching node $v$, that the correct
amount of flow is sent along each edge $(v, u)$, and the condition is written in
terms of $\out(v, i, u)$. Recall that $\out(v, i, u)$ denotes the number of
times that $u$ appears in the first $i$ entries of $\ord(v)$, and observe that
this can be computed in polynomial time, even when $\ord(v)$ is given
succinctly. Thus, the third constraint of a controlled switching flow can also
be checked in polynomial time for succinct games.
\end{proof}

\subsection{\NP-hardness}
\label{subsec:nphardness}

In this section we show that deciding the winner of a one-player 
RSG is \NP-hard. Our construction will produce an explicit RSG, so we obtain \NP-hardness for both explicit and succinct games.
We reduce from 3SAT. Throughout this
section, we will refer to a 3SAT instance with $n$ variables $x_1$, $x_2$, \dots,
$x_n$, and $m$ clauses $C_1$, $C_2$, \dots, $C_m$. Thus, the overall size of the
3SAT formula is $n$ + $m$.
It is well-known~\cite[Thm. 2.1]{Tovey1984} that 3SAT remains \NP-hard even if
all variables
appear in at most three clauses. We make this assumption during our
reduction. 

\paragraph{\bf Overview.} The idea behind the construction is
that the player will be asked to assign values to each variable.
Each variable $x_i$ has a corresponding vertex that will be visited 3
times during the game. Each time this vertex is visited, the player
will be asked to assign a value to $x_i$ in a particular clause $C_j$. If the
player chooses an assignment that \emph{does not} satisfy~$C_j$, then the game
records this by incrementing a counter. If the counter corresponding to any
clause $C_j$ is incremented to three (or two if the clause only has two
variables), then the reachability player immediately loses, since the chosen
assignment fails to satisfy $C_j$.

The problem with the idea presented so far is that there is no mechanism to
ensure that the reachability player chooses a consistent assignment to the
same variable. Since each variable $x_i$ is visited three times, there is
nothing to stop the reachability player from choosing contradictory assignments
to $x_i$ on each visit. To address this, the game also counts how many times
each assignment is chosen for $x_i$. At the end of the game, if the reachability
player has not already lost by failing to satisfy the formula, the game is
configured so that the target is only reachable if the reachability player chose
a consistent assignment.
A high-level overview of the construction for an example formula is given in 
\Cref{fig:overview1}.

\begin{figure}[ht]
\centering
\resizebox{0.7\linewidth}{!}{
\tikzset{box/.style={draw, thick, minimum height=1.8cm,minimum width=0.5cm}}
\tikzstyle{component}=[fill=white, draw, thick, minimum size=0cm,align=center]
\tikzstyle{dest}=[fill=white, thick, minimum size=0cm,align=center]

\begin{tikzpicture}[
	decoration={markings,mark=at position 0.6 with {\arrow{triangle 60}}},
	path/.style={thick},
	clip=false
]

\begin{pgfonlayer}{background}
\node[box, xshift=0cm, yshift=0cm, align=center, text width=1.8cm] (contr) {Controller};
\end{pgfonlayer}{background}


\node[dest,xshift=-1cm] (start) at (contr.west) {start};


\node[dest,xshift=1cm,yshift=0cm] (fail) at (contr.east) {fail};


\node[component,xshift=-3cm,yshift=-1cm] (x1) at (contr.south) {$x_1$};
\node[component,xshift=-1cm,yshift=-1cm]  (x2) at (contr.south) {$x_2$};
\node[component,xshift=1cm,yshift=-1cm]  (x3) at (contr.south) {$x_3$};
\node[component,xshift=3cm,yshift=-1cm]  (x4) at (contr.south) {$x_4$};


\node[dest,xshift=1.5cm,yshift=0cm] (target) at (x4) {target};



\node[component,xshift=0.5cm,yshift=-1cm]  (c1) at (x1.south) {$C_1$};
\node[component,xshift=1cm,yshift=-1cm]  (c2) at (x2.south) {$C_2$};
\node[component,xshift=3.5cm,yshift=-1cm]  (c3) at (x2.south) {$C_3$};

\node[dest,xshift=-1cm,yshift=-1cm]  (x1start) at (x1.south) {start};
\node[dest,xshift=0cm,yshift=-1cm]  (x3start) at (x3.south) {start};
\node[dest,xshift=0.5cm,yshift=-1cm]  (x4start) at (x4.south) {start};
\node[dest,xshift=1.5cm,yshift=-1cm]  (x4start2) at (x4.south) {start};


\node[dest,xshift=-0.5cm,yshift=-1cm]  (c1start) at (c1.south) {start};
\node[dest,xshift=0.5cm,yshift=-1cm]  (c1fail) at (c1.south) {fail};

\node[dest,xshift=-0.5cm,yshift=-1cm]  (c2start) at (c2.south) {start};
\node[dest,xshift=0.5cm,yshift=-1cm]  (c2fail) at (c2.south) {fail};

\node[dest,xshift=-0.5cm,yshift=-1cm]  (c3start) at (c3.south) {start};
\node[dest,xshift=0.5cm,yshift=-1cm]  (c3fail) at (c3.south) {fail};


\draw[path,->] (start) -- (contr);
\draw[path,->] (contr) -- (fail);

\draw[path,->] (contr) -- node[right] {} (x1);
\draw[path,->] (contr) -- node[right] {} (x2);
\draw[path,->] (contr) -- node[right] {} (x3);
\draw[path,->] (contr) -- node[right] {} (x4);

\draw[path,dashed,->] (x1) -- node[right] {} (x2);
\draw[path,dashed,->] (x2) -- node[right] {} (x3);
\draw[path,dashed,->] (x3) -- node[right] {} (x4);
\draw[path,dashed,->] (x4) -- node[right] {} (target);

\draw[path,->] (c1) -- node[right] {} (c1start);
\draw[path,->] (c1) -- node[right] {} (c1fail);
\draw[path,->] (c2) -- node[right] {} (c2start);
\draw[path,->] (c2) -- node[right] {} (c2fail);
\draw[path,->] (c3) -- node[right] {} (c3start);
\draw[path,->] (c3) -- node[right] {} (c3fail);


\draw[path,->] (x1) -- node[right] {} (c1);
\draw[path,->] (x2) -- node[right] {} (c1);

\draw[path,->] (x1) -- node[right] {} (c2);
\draw[path,->] (x2) -- node[right] {} (c2);
\draw[path,->] (x3) -- node[right] {} (c2);

\draw[path,->] (x2) -- node[right] {} (c3);
\draw[path,->] (x3) -- node[right] {} (c3);
\draw[path,->] (x4) -- node[right] {} (c3);

\draw[path,->] (x1) -- node[right] {} (x1start);
\draw[path,->] (x3) -- node[right] {} (x3start);
\draw[path,->] (x4) -- node[right] {} (x4start);
\draw[path,->] (x4) -- node[right] {} (x4start2);

\end{tikzpicture}
}
\caption{\label{fig:overview1} Overview of our construction for one player for the example 
formula $C_1 \land C_2 \land C_3 = (x_1 \lor \lnot x_2) \land (\lnot x_1 \lor \lnot x_2 \lor x_3) \land (x_2 \lor x_3 \lor x_4)$.
Note that the negations of variables in the formula are not relevant for this high-level view; they
will feature in the clause gadgets as explained below. 
The edges for the variable phase are solid, and the edges for the verification phase are dashed.
}
\end{figure}

\begin{figure}[ht]%
\begin{minipage}[b]{0.49\textwidth}%
\begin{center}
\resizebox{0.8\linewidth}{!}{
\begin{tikzpicture}

\node[gadget] (a) {$a^{3n+1}b$};

\node[left of=a, node distance=2.5cm] (in) {start};
\node[below of=a, node distance=1.5cm, switch] (outa) {};
\node[right of=a, node distance=2.5cm] (outb) {fail};

\node[below left=1.25cm and 1.3cm of outa, reach] (x1) {$x_1$};
\node[below left=1.25cm and 0.4cm of outa, reach] (x2) {$x_2$};
\node[below left=1.25cm and -0.5cm of outa, reach] (x3) {$x_3$};
\node[below left=1.25cm and -1.4cm of outa, reach] (x4) {$x_4$};
\node[below left=1.25cm and -2.3cm of outa, reach] (x5) {$x_5$};

\path[->]
    (in) edge (a)
    (a) edge node [left] {$a$} (outa)
    (a) edge node [above] {$b$} (outb)

    (outa) edge [bend right] node [left,xshift=-2mm] {1} (x1)
    (outa) edge [bend right] node [left] {2} (x2)
    (outa) edge [] node [left] {3} (x3)
    (outa) edge [bend left] node [right,xshift=1mm] {4} (x4)
    (outa) edge [bend left] node [right,xshift=2mm] {5} (x5)
    ;

\end{tikzpicture}
}
\caption{The control gadget.}
\label{fig:control}
\end{center}
\end{minipage}
\hfill
\begin{minipage}[b]{0.5\textwidth}
\begin{center}
\resizebox{1\linewidth}{!}{
\begin{tikzpicture}

\node[reach] (a) {$x_i$};

\node[gadget, below left=0.75cm and 0.5cm of a] (t) {$a^3b$};
\node[left of=t, node distance=2.2cm] (tnext) {$x_{i+1}$};
\node[switch, below of=t, node distance=1.75cm] (ts) {};
\node[below left=1.5cm and 1cm of ts.center] (tsi) {$C_i$};
\node[below left=1.5cm and -0.3cm of ts.center] (tsj) {start};
\node[below left=1.5cm and -1.3cm of ts.center] (tsk) {start};

\path[->]
    (t) edge node [left] {a} (ts)
    (t) edge node [above] {b} (tnext)

    (ts) edge [bend right] node [left] {1} (tsi)
    (ts) edge [] node [left] {2} (tsj)
    (ts) edge [bend left] node [left] {3} (tsk)
    ;

\node[gadget, below right=0.75cm and 0.5cm of a] (f) {$a^3b$};
\node[right of=f, node distance=2.2cm] (fnext) {$x_{i+1}$};
\node[switch, below of=f, node distance=1.75cm] (fs) {};
\node[below left=1.5cm and 1cm of fs.center] (fsi) {$C_j$};
\node[below left=1.5cm and -0.3cm of fs.center] (fsj) {$C_k$};
\node[below left=1.5cm and -1.3cm of fs.center] (fsk) {start};
\path[->]
    (f) edge node [left] {a} (fs)
    (f) edge node [above] {b} (fnext)

    (fs) edge [bend right] node [left] {1} (fsi)
    (fs) edge [] node [left] {2} (fsj)
    (fs) edge [bend left] node [left] {3} (fsk)
    ;
\path[->]
    (a) edge [bend right] node [left,xshift=-2mm] {true} (t)
    (a) edge [bend left] node [right,xshift=2mm] {false} (f)
    ;
\end{tikzpicture}
}
\end{center}
\caption{A variable gadget.}
\label{fig:variable}
\end{minipage}
\end{figure}

\paragraph{\bf The control gadget.}
The sequencing in the construction is determined by the control gadget, which is
shown in \Cref{fig:control}. 
Recall that, in our diagramming notation, square vertices
belong to the reachability player, circle vertices are switching nodes, and the
switching order of each switching vertex is labelled on its outgoing edges. Our
diagrams also include \emph{counting gadgets}, which are represented as
non-square rectangles that have labelled output edges. The counting gadget is
labelled by a sequence over these outputs, with the idea being that if the play
repeatedly reaches the gadget, then the corresponding output sequence will be
produced. In \Cref{fig:control} the gadget is labelled by~$a^{3n+1}b$, which means the
first $3n+1$ times the gadget is used the token will be moved along the $a$
edge, and the $3n+2$nd time the gadget is used the token will be moved along the
$b$ edge. This gadget can be easily implemented by a switching node that has
$3n+2$ outgoing edges, the first $3n+1$ of which go to $a$, while the $3n+2$nd
edge goes to $b$. We use gadgets in place of this because it
simplifies our diagrams. 

The control gadget has two phases. In the \emph{variable phase}, 
each variable gadget,
represented by the vertices $x_1$ through $x_n$ is used exactly $3$ times, and
thus overall the gadget will be used $3n$ times. This
is accomplished by a switching node that ensures that each variable is used $3$
times.
After each variable gadget has been visited $3$ times, the control gadget then
sends the token to the $x_1$ variable gadget for the \emph{verification phase}
of the game. In this phase, the reachability player must prove that he gave
consistent assignments to all variables. 
If the control gadget is visited $3n + 2$ times, then the token will be moved to
the \emph{fail} vertex. This vertex has no outgoing edges, and thus is losing
for the reachability player.

\paragraph{\bf The variable gadgets.}
%
%
%
%
%
%
%
%
%
%
%
Each variable $x_i$ is represented by a variable gadget, which is shown in
\Cref{fig:variable}. This gadget will be visited $3$ times in total during
the variable phase, and
each time the reachability player must choose either the true or false edges at
the vertex~$x_i$. In either case, the token will then pass through a counting
gadget, and then move to a switching vertex which either moves the token to a
clause gadget, or back to the start vertex.

It can be seen that the gadget is divided into two almost identical branches.
One corresponds to a true assignment to $x_i$, and the other to a false
assignment to $x_i$. The clause gadgets are divided between the two branches of
the gadget. In particular, a clause appears on a branch if and only if the 
variable appears in that clause and the
choice made by the reachability player \emph{fails} to satisfy the clause. So,
the clauses in which $x_i$ appears positively appear on the false branch of the
gadget, while the clauses in which $x_i$ appears negatively appear on the true
branch. 

The switching vertices each have exactly three outgoing edges. These edges use
an arbitrary order over the clauses assigned to the branch. If there are fewer
than $3$ clauses on a particular branch, the remaining edges of the switching
node go back to the start vertex. Note that this means that a variable can be
involved with fewer than three clauses.

The counting gadgets will be used during the verification phase of the game, in
which the variable player must prove that he has chosen consistent assignments
to each of the variables. Once each variable gadget has been used $3$ times,
the token will be moved to $x_1$ by the control gadget. If the reachability
player has used the same branch three times, then he can choose that branch, and
move to $x_2$, which again has the same property. So, if the reachability player
gives a consistent assignment to all variables, he can eventually move to~$x_n$, and then on to
$x_{n+1}$, which is the target vertex of the game. 
Since, as we will show, there is no other way of reaching $x_{n+1}$, this
ensures that the reachability player must give consistent assignments to the
variables in order to win the game.

\paragraph{The clause gadgets.}
Each clause $C_j$ is represented by a clause gadget, an example of which is
shown in \Cref{fig:clause}.
\begin{figure}[htpb]
\begin{center}
\resizebox{0.4\linewidth}{!}{
\begin{tikzpicture}

\node[gadget] (a) {$a^2b$};

\node[above of=a, node distance=1cm] (in) {};
\node[left of=a, node distance=2.3cm] (outa) {start};
\node[right of=a, node distance=2.3cm] (outb) {fail};

\path[->]
    (in) edge (a)
    (a) edge node [above] {$a$} (outa)
    (a) edge node [above] {$b$} (outb)
    ;

\end{tikzpicture}
}
\end{center}
\caption{A gadget for a clause with three variables.}
\label{fig:clause}
\end{figure}
The gadget counts how many variables have
failed to satisfy the corresponding clause. If the number of times the gadget is
visited is equal to the number of variables involved with the clause, then the
game moves to the fail vertex, and the reachability player immediately loses. In
all other cases, the token moves back to the start vertex.

\paragraph{Correctness.} The following lemma shows that the
reachability player wins the one-player RSG if and only
if the 3SAT instance is satisfiable.

\begin{lem}
\label{lem:one_player_combined}
The reachability player wins the one-player RSG 
if and only if 
the 3SAT instance is satisfiable.
\end{lem}

We split the two directions into two separate lemmas.

\begin{lem}
\label{lem:3SAT_to_rsg_win}
If there is a satisfying assignment to the 3SAT formula, then the reachability
player can win the one-player RSG.
\end{lem}
\begin{proof}
The strategy for the reachability player is as follows: at each variable
vertex $x_i$, choose the branch that corresponds to the value of $x_i$ in the
satisfying assignment. We argue that this is a winning strategy.
First note that the game cannot be lost in a clause gadget during the variable
phase. Since the assignment is satisfying, the play cannot visit a clause gadget
more than twice (or more than once if the clause only has two variables), and therefore the edges from the counting gadgets to the fail
vertex cannot be used. 
Hence, the game will eventually reach the verification phase.  At this
point, since the strategy always chooses the same branch, the play will 
pass through $x_1$, $x_2$, $\dots$, $x_n$, and then arrive at $x_{n+1}$. Since this
is the target, the reachability player wins the game. 
\end{proof}

\begin{lem}
\label{lem:rsg_win_to_3SAT}
If the reachability player wins the one-player RSG, then
there is a satisfying assignment of the 3SAT formula.
\end{lem}
\begin{proof}
We begin by arguing that, if the reachability player wins the game, then he must
have chosen the same branch at every visit to every variable gadget. This holds
because~$x_{n+1}$ can only be reached by ensuring that each variable has a
branch that is visited at least 3 times. The control gadget causes the
reachability player to immediately lose the game if it is visited $3n + 2$
times. Thus, the reachability player must win the game after passing through the
control gadget exactly $3n + 1$ times. The only way to do this is to ensure that
each variable has a branch that is visited exactly 3 times during the variable
phase.

Thus, given a winning strategy for the game, we can extract a consistent
assignment to the variables in the 3SAT instance. Since the game was won, we
know that the game did not end in a clause gadget, and therefore under this
assignment every clause has at least one literal that is true. Thus, the
assignment satisfies the 3SAT instance.
\end{proof}

Note that our game can be written down as an explicit game whose size is
polynomial in $n + m$, so our lower bound applies to both explicit and succinct
games. Hence, we have the following theorem.

\begin{thm}
\label{thm:nphard}
Deciding the winner of an explicit or succinct one-player RSG is \NP-hard.
\end{thm}

\subsection{Memory requirements of winning strategies in one-player games}
\label{sec:memory_oneplayer}

In this section we consider the memory requirements for winning strategies in
one-player reachability switching games. For general history-dependent
strategies, where the player has access to the current switch configuration,
there exist winning strategies that use no memory (this will be argued formally
in the proof of \Cref{lem:memoryupper}). Here we consider the scenario
where the player \emph{does not} have access to the current switch
configuration, but \emph{does} have access to the current vertex and to some
finite memory. We will show an exponential lower bound on the amount of memory
that the player needs to execute a winning strategy.

Consider the game shown in \Cref{fig:memory_one_player}, which 
takes as input a parameter $p$ that we will fix later.
\begin{figure}[htpb]
\begin{center}
\resizebox{0.8\linewidth}{!}{
\begin{tikzpicture}

\node[gadget] (first) {$a^{(p + p^2)}b$};
\node[right of=first, node distance=3cm, draw, reach] (player) {$x$};

\node[gadget, above right=0.6cm and 1.8cm of player] (top) {$a^{p^2}b$};
\node[gadget, below right=0.6cm and 1.8cm of player] (bottom) {$a^{p}b$};

\node[below of=top, node distance=1.35cm] (start2) {start};
\node[left of=first, node distance=2.3cm] (start) {start};
\node[below of=first, node distance=1.5cm] (target) {target};

\node[right of=top, node distance=2.3cm] (topfail) {fail};
\node[right of=bottom, node distance=2.3cm] (bottomfail) {fail};

\path[->]
    (first) edge node [above] {$a$} (player)
    (start) edge (first)
    (first) edge node [left] {$b$} (target)
    (player.east) edge node [left] {top} (top.west)
    (player.east) edge node [left] {bottom} (bottom.west)
    (top.south) edge node [left] {$a$} (start2)
    (bottom.north) edge node [left] {$a$} (start2)
    (top) edge node [above] {$b$} (topfail)
    (bottom) edge node [above] {$b$} (bottomfail)
    ;

\end{tikzpicture}
}
\end{center}
\caption{One-player memory lower bound construction.}
\label{fig:memory_one_player}
\end{figure}

The only control state
for the player is $x$. By construction, $x$
will be visited $p + p^2$ times. Each time, the player must choose either the
top or bottom edge. If the player uses the top edge strictly more than~$p^2$
times, or the bottom edge strictly more than $p$ times, then he will immediately
lose the game. If the player does not lose the game in this way, then after $p^2 + p$
rounds the target will be reached, and the player will win the game.

The player has an obvious winning strategy: use the top edge $p^2$
times and the bottom edge $p$ times. 
Intuitively,
there are two ways that the player could implement the strategy.
(1) Use the bottom edge $p$ times, and then use the top edge $p^2$ times. This
approach uses $p$ memory states to count the number of times the bottom edge has
been used.
(2) Use the bottom edge once, use the top edge $p$ times, and then repeat.
This approach uses $p$ memory states to count the number of times the top state
has been used after each use of the bottom edge.
We can prove that one cannot do significantly better.


\begin{lem}
\label{lem:1pmem}
The reachability player must use at least $p-1$ memory states to win the game
shown in \Cref{fig:memory_one_player}.
\end{lem}

\begin{proof}
Consider a winning strategy and let $M$ denote the number of memory states that it requires.
Run the strategy until the vertex $x$ is visited for the $p$th time, and keep
track of the memory states that are visited by the strategy. At this
point there are two possibilities. If no memory state has been visited twice,
then the strategy has used $p-1$ distinct memory states, and so the claim has
been shown.

Alternatively, if the strategy has used the same memory state twice, then there
must be a cycle of memory states, and since the player has only one vertex, we
know that this cycle will be repeated until the end of the
game. We have the following facts about this cycle:
\begin{itemize}
\item The bottom edge was used used at most $p-1$ times before the cycle
started, but all winning strategies must use the bottom edge $p$ times.
Therefore the cycle must use the bottom edge at least once or the player will
lose.
\item Since the bottom edge cannot be used more than $p$ times, and each
iteration of the cycle uses the bottom edge at least once, it follows that the
cycle cannot be repeated more than $p$ times.
\item The top edge was used at most $p-1$ times before the cycle started, and so
it must be visited at least $p^2 - (p-1)$ times before the game is won. 
\end{itemize}
From the above, we get that each iteration of the cycle
must use the top edge $O(p)$ many times, since otherwise the last two
constraints above could not be satisfied. 
This means that the cycle uses $O(p)$
memory states.


To make the argument above more precise, let $C_T$ and $C_B$ denote the number of times the strategy chose top and
bottom, respectively, in the prefix before we entered the cycle of memory
states. Let $L$ denote the length of the cycle of memory states, and $L_T$
denote the number of memory states on the cycle where the strategy chooses top.
The cycle must use the top edge $p^2 - C_T$ times, and the number of times that
the cycle can repeat is bounded by $p - C_B$. So, we get the following bound on
the number of memory states:
\begin{equation*}
M \ge L \ge L_T \ge \frac{p^2 - C_T}{p - C_B} \ge \frac{p^2 - p}{p} = p-1,
\end{equation*}
which completes the proof.
\end{proof}

We will set $p = 2^{n/2}$ to obtain our lower bound. We show that, even though
$p$ is exponential, it is possible to create an explicit switching gadget that
produces the sequence~$a^{2^{n}}b$ using $n$ switching nodes. 


\begin{figure}
\begin{center}
\begin{tikzpicture}
    \node[state] (one) {$v_1$};
    \node[state, right of=one, node distance=2cm] (two) {$v_2$};
    \node[state, right of=two, node distance=2cm] (three) {$v_3$};
    \node[state, right of=three, node distance=2cm] (four) {$v_4$};

    \node[below of=one, node distance=1.5cm] (aout) {$a$};
    \node[right of=four, node distance=1.5cm] (bout) {$b$};

    \path[->]
        (one) edge node [anchor=north] {2} (two)
        (two) edge node [anchor=north] {2} (three)
        (three) edge node [anchor=north] {2} (four)
        (four) edge node [anchor=north] {2} (bout)
        (one) edge node [anchor=west] {1} (aout)

        (two) edge [bend right=35] node [xshift=0.5cm,yshift=0.1cm] {1} (one)
        (three) edge [bend right=40] node [xshift=0.5cm,yshift=0.2cm] {1} (one)
        (four) edge [bend right=45] node [xshift=0.5cm,yshift=0.2cm] {1} (one)
        ;
\end{tikzpicture}
\end{center}
\caption{A gadget that produces $a^{15}b$. }
\label{fig:gadget}
\end{figure}

To show this, we will build gadgets that produce the sequence $a^xb$
where $x$ is a number encoded in binary. 
The construction is given in the following lemma.
An example gadget produced by the construction is given in \Cref{fig:gadget}.

\begin{lem}
\label{lem:axb}
For all $x \in \nats$ there is an explicit switching gadget of size $\log_2(x)$ with output~$a^xb$.
\end{lem}

\begin{proof}
We will build up the construction recursively. 
Each gadget will have a start state $s$, and two output states $a$ and
$b$. Each time the token enters the gadget at~$s$, it leaves via $a$ or~$b$. 
We
are interested in sequence of outputs generated if the token repeatedly arrives
at $s$. Given a word $w$ over the alphabet $\{a, b\}$ (eg. $abba$) we say that a
gadget produces that word if repeatedly feeding the token through the gadget
produces the sequence~$w$ on the outputs. 
For every $x$, we will denote the
gadget that outputs the word~$a^xb$ as $\gadget(a^xb)$.

For the base case of the recursion, we consider all $x \le 2^0
= 1$. For $x = 0$ we use the following gadget:
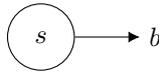
\begin{center}
\begin{tikzpicture}
\node[state] (one) {$s$};

\node[right of=one, node distance=1.5cm] (bout) {$b$};

\path[->]
    (one) edge  (bout)
    ;
\end{tikzpicture}
\end{center}
For $x = 1$ we use the following gadget:
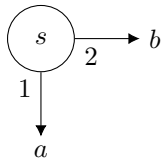
\begin{center}
\begin{tikzpicture}
\node[state] (one) {$s$};

\node[below of=one, node distance=1.5cm] (aout) {$a$};
\node[right of=one, node distance=1.5cm] (bout) {$b$};

\path[->]
    (one) edge node [near start, left] {$1$} (aout) 
    (one) edge node [near start, below] {$2$} (bout)
    ;
\end{tikzpicture}
\end{center}
The correctness of both of these gadgets is self-evident.

Now, suppose that we have gadgets for all $x' \le 2^{i-1}$, and let $x$ be a
number with $2^{i-1} < x \le 2^i$. If $x$ is odd, then we use the following
construction:
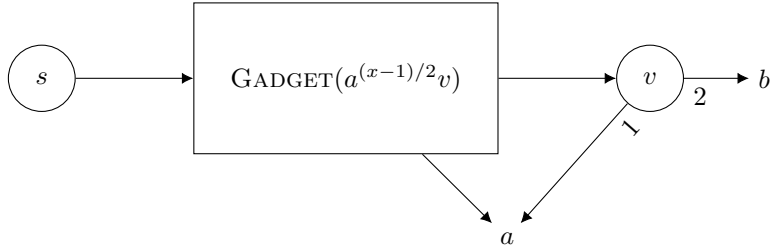
\begin{center}
\begin{tikzpicture}
\node[state] (one) {$s$};
\node[draw, minimum width=4cm, minimum height=2cm, right of=one, node distance=4cm] (gadget) {$\gadget(a^{(x-1)/2}v)$};
\node[state, right of=gadget, node distance=4cm] (two) {$v$};

\node[below right of=gadget, node distance=3cm] (aout) {$a$};
\node[right of=two, node distance=1.5cm] (bout) {$b$};

\path[->]
    (one) edge (gadget)
    (gadget) edge (aout)
    (gadget) edge (two)
    (two) edge node [near start, below, sloped] {$2$} (bout)
    (two) edge node [pos=0.1, below, sloped] {$1$} (aout)
    ;
\end{tikzpicture}
\end{center}
The gadget produces the sequence $a^{(x-1)/2} \cdot a \cdot a^{(x-1)/2} \cdot b
= a^x b$, where $\cdot$ denotes concatenation.

If $x$ is even, then we use the following gadget.
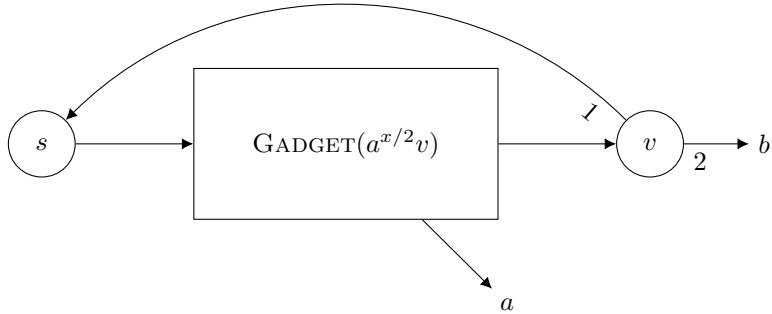
\begin{center}
\begin{tikzpicture}
\node[state] (one) {$s$};
\node[draw, minimum width=4cm, minimum height=2cm, right of=one, node distance=4cm] (gadget) {$\gadget(a^{x/2}v)$};
\node[state, right of=gadget, node distance=4cm] (two) {$v$};

\node[below right of=gadget, node distance=3cm] (aout) {$a$};
\node[right of=two, node distance=1.5cm] (bout) {$b$};

\path[->]
    (one) edge (gadget)
    (gadget) edge (aout)
    (gadget) edge (two)
    (two) edge node [near start, below, sloped] {$2$} (bout)
    (two) edge [bend right=45] node [pos=0.05, below, sloped] {$1$} (one)
    ;
\end{tikzpicture}
\end{center}
The gadget produces $a^{x/2} \cdot a^{x/2} \cdot b = a^x b$. 

So, we have provided a family of gadgets the produce sequences of the form
$a^xb$. Since each iteration of the recursion divides $x$ by two, and since each
iteration adds at most one new state, we have that $\gadget(a^xb)$ uses
$\log_2(x)$ switching nodes. 
\end{proof}

\begin{thm}
\label{thm:memone}
The number of memory states needed in an explicit one-player reachability
switching game is $2^{\Omega{(n)}}$, where $n$ is the number of states.
\end{thm}

\begin{proof}
If we set $p = 2^{k/2}$, and use the
gadgets from \Cref{lem:axb}, 
then the game in \Cref{fig:memory_one_player} has $n = O(\log (p^2)) = O(\log (2^n)) = O(k)$ states. 
\Cref{lem:1pmem} shows that the reachability player needs $p - 1 =
2^{k/2} - 1$ memory states to win the game. Hence, $2^{\Omega{(n)}}$ memory
states are required.
\end{proof}

\section{Two-player reachability switching games}
\label{sec:twoplayer}

\subsection{Containment in \EXPTIME}
\label{sec:2pcontain}
We first observe that solving a two-player RSG lies in
\EXPTIME. This can be proved easily, either by blowing the game up into an
exponentially sized reachability game, or equivalently, by simulating the game on
an alternating polynomial-space Turing machine. 

\begin{thm}
\label{thm:2player-upper}
Deciding the winner of an RSG is in \EXPTIME.
\end{thm}

\begin{proof}
We prove this by showing that the game can be simulated by an \emph{alternating
Turing machine}, which is a machine that has both existential and
universal non-determinism. It has been shown that
\APSPACE=~\EXPTIME~\cite{CKS81}, which means that if we can devise an algorithm
that runs in polynomial space on an alternating Turing machine, then we can
obtain an algorithm that runs in exponential time on a deterministic Turing
machine.

It is straightforward to implement an explicit or succinct 
RSG on an alternating Turing machine. The machine simulates a run of
the game. It starts by placing a token on the starting state. It then simulates
each step of the game. When the token arrives at a vertex belonging to the
reachability player, it uses existential non-determinism to choose a move for
that player. When the token arrives at a vertex belonging to the safety player,
it uses universal non-determinism to choose a move for that player. The moves at
the switching nodes are simulated by remembering the current switch
configuration, which can be done in polynomial space. The machine accepts if and
only if the game arrives at the target state.

This machine uses polynomial space, because it needs to remember the switch
configuration. Note that it still uses polynomial space even for succinct games,
since a switch configuration for a succinct game can be written down as a list
of $n$ numbers expressed in binary. This completes the proof of \Cref{thm:2player-upper}.
\end{proof}

\subsection{\PSPACE-hardness}
We show that deciding the winner of an explicit two-player 
RSG is \PSPACE-hard, by reducing \emph{true quantified boolean
formula} (TQBF), the canonical \PSPACE-complete problem, to our problem.
Throughout this section we will refer to a TQBF instance 
$\exists x_1 \forall x_2 \dots \exists x_{n-1} \forall x_n \cdot \phi(x_1, x_2, \dots,
x_n),$
where $\phi$ denotes a boolean formula given in negation normal form, which
requires that negations are only applied to variables, and not sub-formulas.
The problem is to decide whether this formula is true.

\paragraph{\bf Overview.}
We will implement the TQBF formula as a game between the reachability player and
the safety player. This game will have two phases. In the \emph{quantifier
phase}, the two players assign values to their variables in the order specified
by the quantifiers. In the \emph{formula phase}, the two players determine
whether $\phi$ is satisfied by these assignments by playing the standard
model-checking game for propositional logic. The target state of the game is
reached if and only if the model checking game determines that the formula is satisfied. 
This high-level view of our construction is depicted in \Cref{fig:overview2}.
\begin{figure}[htpb]
\centering
\resizebox{0.65\linewidth}{!}{
\tikzset{box/.style={draw, thick, minimum height=1.8cm,minimum width=0.5cm}}
\tikzstyle{component}=[fill=white, draw, thick, minimum size=0cm, align=center]
\tikzstyle{dest}=[fill=white, minimum size=0.5cm]

\begin{tikzpicture}[
	decoration={markings,mark=at position 0.6 with {\arrow{triangle 60}}},
	path/.style={thick},
	clip=false
]

\begin{pgfonlayer}{background}
\node[box, xshift=0cm, yshift=0cm, align=center, text width=2cm] (formula) {Formula $\phi$};
\end{pgfonlayer}{background}


\node[component,xshift=-3cm,yshift=-0.75cm]  (x1) at (formula.south) {$x_1$};
\node[component,xshift=1cm,yshift=-0.75cm]  (x2) at (formula.south) {$x_2$};
\node[xshift=-1cm,yshift=-0.75cm]  (bphantom) at (formula.south) {$\cdots$};
\node[component,xshift=3cm,yshift=-0.75cm]  (xn) at (formula.south) {$x_n$};


\node[dest,xshift=-1.5cm,yshift=0cm]  (start) at (x1) {start};


\node[dest,xshift=-0.5cm,yshift=-1cm]  (t1) at (x1.south) {target};
\node[dest,xshift=0.5cm,yshift=-1cm]  (f1) at (x1.south) {fail};
\node[dest,xshift=-0.5cm,yshift=-1cm]  (t2) at (x2.south) {target};
\node[dest,xshift=0.5cm,yshift=-1cm]  (f2) at (x2.south) {fail};
\node[dest,xshift=-0.5cm,yshift=-1cm]  (tn) at (xn.south) {target};
\node[dest,xshift=0.5cm,yshift=-1cm]  (fn) at (xn.south) {fail};



\draw[path,dashed,->] (start) -- (x1);

\draw[path,dashed,->] (x2) -- node[right] {} (xn);
\draw[path,dashed,->] (bphantom) -- node[right] {} (x2);
\draw[path,dashed,->] (x1) -- node[right] {} (bphantom);


\draw[path,dotted,->] (xn.north) to[out=-270, in=45] (formula.north);

\draw[path,<-] (xn) -- node[right] {} (formula);
\draw[path,<-] (x2) -- node[right] {} (formula);
\draw[path,<-] (x1) -- node[right] {} (formula);
\draw[path,->] (x1) -- (t1);
\draw[path,->] (x1) -- (f1);
\draw[path,->] (x2) -- (t2);
\draw[path,->] (x2) -- (f2);
\draw[path,->] (xn) -- (tn);
\draw[path,->] (xn) -- (fn);

\end{tikzpicture}
}
\caption{\label{fig:overview2} High-level overview of our construction for two players.
	The dashed lines between variables are part of the first, quantifier phase;
	the dotted line from variable $x_n$ to the Formula is the transition between
	phases, and the solid edges are part of the second, formula phase.} 
\end{figure}

\paragraph{\bf The quantifier phase.}
Each variable in the TQBF formula will be represented by an \emph{initialization
gadget}. The initialization gadget for an existentially quantified variable is
shown in \Cref{fig:2pvar}.
\begin{figure}[htpb]
\begin{minipage}[b]{0.48\linewidth}
\begin{center}
\resizebox{1\linewidth}{!}{
\begin{tikzpicture}

\node [reach] (d) {$d_i$};
\node [switch, below left=0.5cm and 0.8cm of d] (pos) {$x_i$};
\node [switch, below right=0.5cm and 0.8cm of d] (neg) {$\lnot x_i$};
\node [switch, below=2cm of d] (f) {$f_i$};

\node [above=0.5cm of d] (in) {from $x_{i-1}$};
\node [left=0.8cm of pos] (pt) {target};
\node [right=0.8cm of neg] (nt) {target};
\node [right=1.5cm of f] (out) {to $x_{i+1}$};
\node [left=1.5cm of f] (fail) {fail};

\path[->]
    (d) edge [bend right] (pos)
    (d) edge [bend left] (neg)
    (pos) edge node [left] {1} (f)
    (neg) edge node [right] {1} (f)

    (pos) edge node [above] {2} (pt)
    (neg) edge node [above] {2} (nt)

    (in) edge (d)
    (f) edge node [below] {1} (out)
    (f) edge node [below] {2} (fail)

    ;

\end{tikzpicture}
}
\end{center}
\caption{The initialization gadget for an existentially quantified variable $x_i$.}
\label{fig:2pvar}
\end{minipage}
\hfill
\begin{minipage}[b]{0.48\textwidth}
\begin{center}
\resizebox{0.65\linewidth}{!}{
\begin{tikzpicture}

\node[safety] (a) {$\land_1$};
\node[above=0.5cm of a] (in) {from $f_{n}$};

\node[reach, below left=0.5cm and 0.75cm of a] (b) {$\lor_1$};
\node[safety, below right=0.7cm and 0.75cm of a] (c) {$\land_2$};

\node[below left=0.5cm and 0mm of b] (x1) {$x_1$};
\node[below right=0.5cm and 0mm of b] (x2) {$\lnot x_2$};
\node[below left=0.5cm and 0mm of c] (x3) {$\lnot x_3$};
\node[below right=0.5cm and 0mm of c] (x4) {$x_4$};

\path[->]
    (in) edge (a)
    (a) edge (b)
    (a) edge (c)
    (b) edge (x1)
    (b) edge (x2)
    (c) edge (x3)
    (c) edge (x4)
    ;
\end{tikzpicture}
}
\end{center}
\caption{The formula phase game for the formula $(x_1 \lor \lnot x_2) \land
\lnot x_3 \land x_4$.}
\label{fig:2pmod}
\end{minipage}
\end{figure}

The gadget for a universally quantified
variable is almost identical, but the state~$d_i$ is instead controlled by the
safety player.

During the quantifier phase, the game will start at $d_1$, and then pass through
the gadgets for each of the variables in sequence. In each gadget, the
controller of $d_i$ must move to either~$x_i$ or $\lnot x_i$. In either case,
the corresponding switching node moves the token to $f_i$, which then
subsequently moves the token on to the gadget for $x_{i+1}$.

The important property to note here is that once the player has made a choice,
any subsequent visit to $x_i$ or $\lnot x_i$ will end the game. Suppose that the
controller of $d_i$ chooses to move to $x_i$. If the token ever arrives at $x_i$
a second time, then the switching node will move to the target vertex and the
reachability player will immediately win the game. If the token ever arrives at
$\lnot x_i$ the token will move to $f_i$ and then on to the fail vertex, and the
safety player will immediately win the game. The same property holds
symmetrically if the controller of $d_i$ chooses $\lnot x_i$ instead. 
In this way, the controller of $d_i$ selects an assignment to~$x_i$. Hence, the
reachability player assigns values to the existentially quantified variables,
and the safety player assigns values to the universally quantified variables.

\paragraph{\bf The formula phase.}
%
%
%
%
%
%
Once the quantifier phase has ended, the game moves into the formula phase.
In this phase the two players play a game to determine whether $\phi$ is
satisfied by the assignments to the variables. This is the standard model
checking game for first order logic. The players play a game on the parse tree
of the formula, starting from the root. The reachability player controls the
$\lor$ nodes, while the safety player controls the $\land$ nodes (recall that
the game is in negation normal form, so there are no internal $\lnot$ nodes.)
Each leaf is either a variable or its negation, which in our game are
represented by the $x_i$ and~$\lnot x_i$ nodes in the initialization gadgets. An
example of this game is shown in \Cref{fig:2pmod}. In our diagramming
notation, nodes controlled by the safety player are represented by triangles.

Intuitively, if $\phi$ is satisfied by the assignment to $x_1, \ldots, x_n$,
then no matter what the safety player does, the reachability player is 
able to reach a leaf node corresponding to a true assignment, and as 
mentioned earlier, he will then immediately win the game. Conversely, if~$\phi$
is not satisfied, then no matter what the reachability player
does, the safety player can reach a leaf corresponding to a false assignment,
and then immediately win the game.


\begin{lem}
\label{lem:QBFtrue}
The reachability player wins the RSG if and only if the QBF formula is true.
\end{lem}

\begin{proof}
If the QBF formula is true, then during the quantifier phase, no matter what 
assignments the safety player
picks for the universally quantified variables, the reachability player can
choose values for the existentially quantified variables in order to make $\phi$
true. Then, in the formula phase the reachability player has a strategy to
ensure that he wins the game, by moving to a node $x_i$ or~$\lnot x_i$ that was
used during the quantifier phase. 

Conversely, and symmetrically, if the QBF formula is false then the safety
player can ensure that the assignment does not satisfy $\phi$ during the
quantifier phase, and then ensure that the game moves to a node $x_i$ or $\lnot
x_i$ that was not used during the quantifier phase. This ensures that the safety
player wins the game.
\end{proof}

Since we have shown the lower bound for explicit games, we also get the same
lower bound for succinct games as well. We have shown the following theorem.

\begin{thm}
\label{thm:2p}
Deciding the winner of an explicit or succinct RSG is
\PSPACE-hard.
\end{thm}

Note that all runs of the game have polynomial length, a property that is not
shared by all RSGs. This gives us the following
corollary.

\begin{cor}
\label{cor:pspacecomplete}
Deciding the winner of a polynomial-length RSG is
\PSPACE-complete.
\end{cor}

\begin{proof}
Hardness follows from \Cref{thm:2p}. For containment, observe that the
simulation by an alternating Turing machine described in
\Cref{sec:2pcontain} runs in polynomial time whenever the game terminates
after a polynomial number of steps. Hence, we can use the fact
that \AP=~\PSPACE~\cite{CKS81} to obtain a deterministic
polynomial space algorithm for solving the problem.
\end{proof}

\subsection{Memory requirements for two-player games}
\label{sec:memory}
\begin{figure}[htpb]
\begin{center}
\resizebox{0.65\linewidth}{!}{
\begin{tikzpicture}
\node [safety] (x) {$x$};

\node [reach, below=1.8cm of x] (y) {$y$};

\node [switch, right=1.8cm of x] (a) {$a$};
\node [switch, below=1.6cm of a] (b) {$b$};

\node [switch, below right=0.6cm and 1.5cm of a] (c) {$c$};

\node [below=0.5cm of a] (t) {target};

\node [right=1cm of c] (cf) {fail};
\node [above=1cm of c] (cy) {y};

\node [left=1cm of x] (start) {start};

\path[->]
    (x) edge (a)
    (x) edge (b)
    (y) edge (a)
    (y) edge (b)

    (a) edge node [above] {1} (c)
    (b) edge node [below] {1} (c)

    (a) edge node [right] {2} (t)
    (b) edge node [right] {2} (t)

    (c) edge node [above] {2} (cf)
    (c) edge node [left] {1} (cy)

    (start) edge (x)
    ;
\end{tikzpicture}
}
\end{center}
\caption{An RSG in which the reachability player needs to use memory.}
\label{fig:memory}
\end{figure}
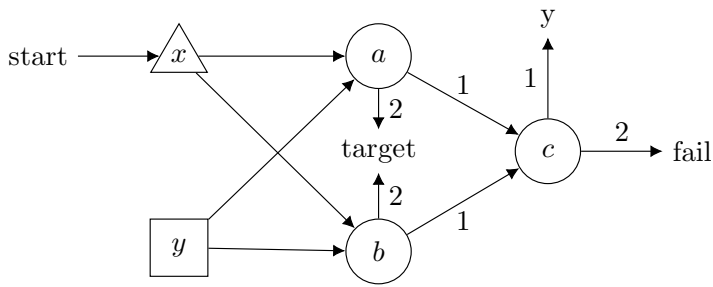

%
%
As in \Cref{sec:memory_oneplayer}, we now consider the scenario in which
the player does not have access to the switch configuration, but does have
access to a finite memory.
We are able to show
a stronger memory lower bound for two-player games compared to one-player games.
\Cref{fig:memory} shows a simple gadget that forces the reachability player to use
memory.
The game starts by
allowing the safety player to move the token from $x$ to either $a$ or $b$. Whatever the choice,
the token then moves to~$c$ and then on to $y$. At this point, if the
reachability player moves the token to the node chosen by the safety player, then the token will arrive 
at the target node and the reachability player will win. If the
reachability player moves to the other node, the token
will move to $c$ for a second time, and then on to the fail vertex, which is
losing for the reachability player. Thus, every winning strategy of the
reachability player must remember the choice made by the safety player.

We can create a similar gadget that forces the safety player to
use memory by swapping the players. In the modified gadget, the safety
player has to choose the vertex \emph{not chosen} by the reachability
player. Thus, in an RSG, winning strategies for both
players need to use memory.
By using $k$ copies of the memory gadget, we can show the following lower
bound. Observe that, in contrast to \Cref{thm:memone}, no asymptotics are
used in this lower bound.

\begin{lem}
\label{lem:memorylower}
In an explicit or succinct RSG, winning strategies for both players may need
to use~$2^{n/6}$ memory states, where $n$ is the number of states.
\end{lem}

\begin{proof}
Consider a game with $k$ copies of the memory gadget shown in
\Cref{fig:memory}, but modified so that the following sequence of events
occurs:
\begin{enumerate}
\item The safety player selects $a$ or $b$ in all gadgets, one at a time.
\item The safety player then moves the game to one of the $y$ vertices in one of
the gadgets.
\item The reachability player selects $a$ or $b$ as normal, and then either wins
or loses the game.
\end{enumerate}
Each gadget has five states giving $5 \cdot k$ states in total, and one extra
state is needed for the safety player to make the decision in Step 2. Hence, the
total number of states is $n = 5 \cdot k + 1 \le 6 \cdot k$, since we can assume
that $k \ge 1$. 

The reachability player has an obvious winning strategy in this game, which is
to remember the choices that the safety player made, and choose the same
vertex in the third step. As the safety player makes $k$ binary decisions,
this strategy uses $2^k$ memory states.

On the other hand, if the reachability player uses a strategy $\sigma$ with
strictly less than $2^k$ memory states, then the safety player can win the game in the following
way. There are $2^k$ different switch configurations that the safety player can
create at the end of the first step of the game. By the pigeon-hole principle
there exists two distinct configurations $C_1$ and $C_2$ that are mapped to the
same memory state by $\sigma$. The safety player selects a gadget $i$ that
differs between $C_1$ and $C_2$, and determines whether $\sigma$ selects $a$ or $b$ for
gadget $i$. He then selects the configuration that that is consistent with the
other option, so if $\sigma$ chooses $a$ the safety player chooses the
configuration $C_j$ that selects $b$. He then sets the gadgets according to
$C_j$ in step 1, and moves the game to gadget $i$ in step 2. The reachability
player will then select the vertex not chosen in step 1, so he loses the game.

Hence, the reachability player must use $2^k$ memory states to win the game.
Since the game has $n \le 6\cdot k$ states, this gives a lower bound of
$2^{n/6}$ memory states.
Finally, observe that we can obtain the same lower bound for the safety player
by swapping the roles of both players in this game. 
\end{proof}

\paragraph{\bf Straightforward upper bound.} 
It should be noted that
that exponential
memory is sufficient in a two-player reachability switching game. We say that a
strategy is a \emph{switch configuration strategy} if it simply remembers the
current switch configuration. Any such strategy uses at most exponentially many
memory states. For games with binary switch nodes, these strategies use exactly
$2^n$ memory states, where $n$ is the number of switching nodes.

\begin{lem}
\label{lem:memoryupper}
In a reachability switching game, both players have winning switch configuration
strategies.
\end{lem}

\begin{proof}
Let $G = (V, E, \vr, \vs, \vswi, o, s, t)$ be a reachability switching game, and let
$\mathcal{C}$ denote the set of all switch configurations in this game. 
Consider the ``blown-up'' reachability game~$G'$ played on $V \times \mathcal{C}$, where there are
no switching nodes, but instead the successor of a vertex $(v, C)$ with $v \in
\vswi$ is determined by $C$. It is straightforward to show that the reachability
player wins the game $G'$ if and only if he wins the
original game. Both players in a reachability game have positional winning
strategies. Therefore, if a player can win in $G'$, then he can also win in $G$
using a switch configuration strategy that always plays according to the
positional winning strategy in~$G'$.
\end{proof}

\section{Zero-player reachability switching games}
\label{sec:zeroplayer}

In this section we consider zero-player RSGs, i.e., where $\vr = \vs = \emptyset$. 

\subsection{Explicit zero-player games}

We show that deciding the winner of an explicit zero-player game is \NL-hard. To
do this, we reduce from the problem of deciding $s$-$t$ connectivity in a
directed graph.
The idea is to make every node in the graph a switching node. We then begin a
walk from $s$. If, after $|V|$ steps we have not arrived at $t$, we go back to
$s$ and start again. So, if there is a path from $s$ to $t$,
then the switching nodes must eventually send the token along that path.

Formally, given a graph $(V, E)$, we produce a zero-player 
RSG played on $V \times V \cup \{\textsf{fin}\}$, where the second component of each state
is a counter that counts up to $|V|$, as follows:
\begin{itemize}
\item The start vertex is $(s, 1)$. 
\item The target vertex is $\textsf{fin}$. 
\item For all $k$, the vertex $(t, k)$ has a single outgoing edge to
$\textsf{fin}$. This means that if we ever reach~$t$ then we win the game.
\item Each vertex $(v, k)$ with $v \ne t$ and $k < |V|$ has the following outgoing edges.
\begin{itemize}
\item If $v$ has at least one outgoing edge, then $(v, k)$ has an edge to $(u,
k+1)$, for each edge $(v, u) \in E$.
\item If $v$ has no outgoing edges, then $(v, k)$ has a single outgoing edge to
$(s, 1)$.
\end{itemize}
In other words, the switching node $(v, k)$ cycles between the outgoing edges of
$v$, and increases the count by $1$, or it goes back to the start vertex if $v$ is a dead-end.
\item Each vertex $(v, |V|)$ with $v \ne t$ has 
a single outgoing edge to $(s, 1)$. This means that when the count reaches $|V|$
the we start again from $(s, 1)$.
\end{itemize}
This game can be constructed in logarithmic space by looping over each element
in $V \times V \cup \{\textsf{fin}\}$ and producing the correct outgoing edges.


\begin{thm}
\label{thm:NLhard}
Deciding the winner of an explicit zero-player RSG is
\NL-hard under logspace reductions.
\end{thm}

\begin{proof}

We must argue that there is a path from $s$ to $t$ if and only if the
zero-player reachability game eventually arrives at $\textsf{fin}$. By definition,
if the game arrives at \text{fin}, then there must be a path from $s$ to $t$,
since all paths from $(s, 1)$ to a vertex $(t, k)$ only use edges from the
original graph.

For the other direction, suppose that there is a path from $s$ to $t$, but the
game never arrives at $\textsf{fin}$. By construction, if the game does not reach
$\textsf{fin}$, then $(s,1)$ is visited infinitely often. Since $(s, 1)$ is a
switching state, we can then argue that the vertex $(v, 2)$ is visited
infinitely often for every successor $v$ of $s$. Carrying on this argument
inductively allows us to conclude that if there is a path of length $k$ from $s$
to $v$, then the vertex $(v, k)$ is visited infinitely often, which provides our
contradiction. 
\end{proof}

\subsection{Succinct games}
\label{sec:succinct}

Deciding reachability for succinct zero-player games still lies
in \NP~$\cap$~\coNP. This can be shown using essentially the same arguments that were used to show
\NP~$\cap$~\coNP containment for explicit games~\cite{DGKMW17}.
The fact that the problem lies in \NP follows from
\Cref{thm:np}, since every succinct zero-player game is also a succinct
one-player game, and so a switching flow can be used to witness
reachability. To put the problem in \coNP, one can follow the original proof
given by Dohrau et al.~\cite[Theorem 3]{DGKMW17} for explicit games. This proof
condenses all losing and infinite plays into a single failure state, and
then uses a switching flow to witness reachability for that failure state. Their
transformation uses only the graph structure of the game, and not the switching
order, and so it can equally well be applied to succinct games.

In contrast to explicit games, we can show a stronger lower bound of \P-hardness
for succinct games. We will reduce from the problem of evaluating a boolean
circuit (the \emph{circuit value problem}), which is one of the canonical
\P-complete problems. 
We will assume that the circuit has fan-in and fan-out 2,
that all gates are either AND-gates or OR-gates, and that the circuit is
\emph{synchronous}, meaning that the outputs of the circuit have depth $1$, and
all gates at depth $i$ get their inputs from gates of depth exactly $i+1$.
This is Problem A.1.6
``Fanin~2, Fanout 2 Synchronous Alternating Monotone CVP''
of Greenlaw et al.~\cite{GHR95}. We will reduce from the following decision problem: for a given input
bit-string $B \in \{0, 1\}^n$, and a given output gate $g$, is $g$ evaluated to
true when the circuit is evaluated on $B$?

\paragraph{\bf Boolean gates.}
We will simulate the gates of the circuit using switching nodes. 
A gate at depth $i > 1$ is connected to exactly two gates of depth $i+1$ from
which it gets its inputs, and exactly two gates at depth $i-1$ to which it sends
its output. If a gate evaluates to true, then it will send a \emph{signal} to
the output-gates, by sending the token to that gate's gadget. More precisely,
for a gate of depth $i > 1$, the following signals are sent.
If the gate evaluates to true, then the gate will send the token exactly
$2^{i-1}$ times to each output gate. 
If the gate evaluates to false, then the gate will send the token exactly
$0$ times to each output gate.
So the number of signals sent by a gate grows exponentially with the depth of
that gate. 

\Cref{fig:zero_and} shows our construction for an AND-gate of depth $2$.
\begin{figure}[htpb]
\begin{minipage}[t]{0.45\textwidth}
\begin{center}
\resizebox{0.75\linewidth}{!}{
\begin{tikzpicture}

\node[node distance=3cm, draw, switch] (switch) {$x$};

\node[above left=0.5cm and 1.8cm of switch] (in1) {$I_1$};
\node[below left=0.5cm and 1.8cm of switch] (in2) {$I_2$};
\node[above=1.2cm of switch] (start) {control};

\node[above right=0.5cm and 1.8cm of switch] (out1) {$O_1$};
\node[below right=0.5cm and 1.8cm of switch] (out2) {$O_2$};

\path[->]
    (in1.east) edge ([yshift=5pt] switch.west)
    (in2.east) edge ([yshift=-5pt] switch.west)
	(switch.north) edge node [left] {$1,2,3,4$} (start.south)
	([yshift=5pt] switch.east) edge node [below,xshift=0.1cm] {$5,6$} (out1.west)
	([yshift=-5pt] switch.east) edge node [above,xshift=0.1cm] {$7,8$} (out2.west)
    ;

\end{tikzpicture}
}
\end{center}
\caption{An AND-gate of depth $2$.}
\label{fig:zero_and}
\end{minipage}
\hfill
\begin{minipage}[t]{0.45\textwidth}
\begin{center}
\resizebox{0.75\linewidth}{!}{
\begin{tikzpicture}

\node[node distance=3cm, draw, switch] (switch) {$y$};

\node[above left=0.5cm and 1.8cm of switch] (in1) {$I_1$};
\node[below left=0.5cm and 1.8cm of switch] (in2) {$I_2$};
\node[above=1.2cm of switch] (start) {control};

\node[above right=0.5cm and 1.8cm of switch] (out1) {$O_1$};
\node[below right=0.5cm and 1.8cm of switch] (out2) {$O_2$};

\path[->]
    (in1.east) edge ([yshift=5pt] switch.west)
    (in2.east) edge ([yshift=-5pt] switch.west)
	(switch.north) edge node [left] {$5,6,7,8$} (start.south)
	([yshift=5pt] switch.east) edge node [below,xshift=0.1cm] {$1,2$} (out1.west)
	([yshift=-5pt] switch.east) edge node [above,xshift=0.1cm] {$3,4$} (out2.west)
    ;

\end{tikzpicture}
}
\end{center}
\caption{An OR-gate of depth $2$.}
\label{fig:zero_or}
\end{minipage}
\end{figure}
It consists of a single switching node (with a succinct order). Further, $I_1$ and $I_2$
are two input edges that come from the two inputs to this gate, and $O_1$ and
$O_2$ are two output edges that go to the outputs of this gate. The control
state is a special state that drives the construction, which will be described
later.
The switching order was generated by the following rules. For a gate at depth $i$, the switching order of an AND-gate is
defined so that
the first $2^i$ positions in the switching order go to control,
the next~$2^{i-1}$ positions in the switching order go to $O_1$,
and the final~$2^{i-1}$ positions in the switching order go to~$O_2$.
Observe that this switching order captures the behavior of an AND-gate. If the
gadget receives~$2^i$ signals from both inputs, then it sends~$2^{i-1}$ signals to both outputs. On the other hand, if at least one of the two
inputs sends no signals, then the gadget sends no signals to the outputs.


The same idea is used to implement OR-gates. \Cref{fig:zero_or} shows the
construction for an OR-gate of depth $2$. For an OR-gate of depth $i$ we have
that
the first $2^{i-1}$ positions in the switching order go to $O_1$,
the next $2^{i-1}$ positions in the switching order go to $O_2$,
and the final $2^i$ positions in the switching order go to control.
These conditions simulate an OR-gate. If either of the inputs produces $2^i$
input signals, then~$2^{i-1}$ signals are sent to both outputs. If both inputs
produce no signals, then no signals are sent to either output.

\paragraph{\bf The control state and the depth~$1$ gates.}
The purpose of the control state is to send the correct signals to the input
gates. Specifically, the input bits that are true should receive~$2^d$ signals,
where $d$ is the depth of the circuit, while the input bits that are false
should receive no signals. 

This is achieved in the following way. The control state is a single switching
node that has the following switching order.
\begin{itemize}
\item Each input edge to a gate at depth $d$ refers to some bit contained in the
bit-string $B$. The control state sends~$2^d$
inputs using that edge if that bit is true, and $0$ inputs
using that edge if that bit is false.
\item Once those signals have been sent, the control state moves the token to an
absorbing failure state.
\end{itemize}
The token begins at the control state.

Each gate at depth $1$ is
represented by a single state, and has the same structure and switch
configuration as the gates at depth $i > 1$. The only difference is the
destination of the edges $O_1$ and~$O_2$. 
The gate $g$ (which we must evaluate) sends all outputs to an absorbing target state.
All other gates send all outputs back to the control state.

\begin{lem}
\label{lem:succinct}
The token reaches the target state if and only if the gate $g$ evaluates to true
when the circuit is evaluated on the bit-string $B$.
\end{lem}

\begin{proof}
We prove the two directions separately.

\paragraph{\bf The $\Rightarrow$ direction.}
Here we must show that if the game reaches the target state, then gate $g$
evaluates to true on the input bit-string. We prove the following two statements
by induction:
\begin{enumerate}
\item Every gate at depth $i$ receives at most $2^{i}$ signals from each of its
inputs.
\item For all gates $g'$ in the circuit, if $g'$ sends at least one signal to
one of its outputs, then $g'$ evaluates to true when the circuit is evaluated on
$B$. 
\end{enumerate}
Note that this is sufficient to prove the claim, since point (2) implies that if
gate $g$ sends the token to the target, then $g$ must evaluate to true when the
circuit is evaluated on $B$.

For the base case, we use the signals generated by the control state. Note that
the control state produces either $2^d$ inputs or $0$ inputs for each input
line, which proves point (1), and the control state sends signals to a gate of
depth $d$ if and only if the corresponding bit of $B$ is true, which proves
point (2).

For the inductive step, let $g'$ be a gate at depth $i$, and suppose that $g'$
is an AND-gate. To prove point (1) note that, by the inductive hypothesis, the
gate $g'$ can receive at most~$2^{i}$ input signals from each input, giving a
total of $2^{i+1}$ input signals in total. No matter whether~$g'$ is
an AND-gate or an OR-gate, our gadgets ensure that at most $2^{i-1}$ signals can be sent to each output,
and any remaining signals will be sent to the control state.

For point (2), let us first assume that $g'$ is an AND-gate. If $g'$ sends a
signal to one of its outputs, then it must have received strictly more than
$2^i$ input signals. Point (1) tells us that the only way this is possible is if
both of input gates sent signals to $g'$. Thus, by the inductive hypothesis,
both of the inputs to $g'$ evaluate to true when the circuit is evaluated on
$B$, and therefore $g'$ must also evaluate to true when the circuit is evaluated
on $B$.

Note suppose that $g'$ is an OR-gate. By construction, if $g'$ sends a signal to one of
its outputs, then at least one of the inputs to $g'$ must have sent a signal to $g'$. By
point (2) of the inductive hypothesis, this means that at least one input of
$g'$ evaluates to true when the circuit is evaluated on $B$. This means that
$g'$ must also evaluate to true when the circuit is evaluated on $B$.

\paragraph{\bf The $\Leftarrow$ direction.}
We show that if gate $g$ evaluates to true on the bit-string $B$, then the
target state will be reached. So, suppose for the sake of contradiction, that
$g$ evaluates to true, but the target state was not reached. Note that the game
cannot continue indefinitely, because the control state appears on every cycle
of the game, and eventually the control state will send the token to the
absorbing failure state. So, since the target was not reached, this means that
the token must have arrived at the failure state.

Since the token arrived at the failure state, this means that the gates at depth
$d$ received the correct input signals for the bit-string $B$. By construction,
this means that they outputted correct signals to the gates at depth $d-1$.
Applying this reasoning inductively, we can conclude that the gate $g$ received
correct input signals from its inputs. But, since gate $g$ evaluates to true,
this means that it sent the token to the absorbing target state, which
contradicts the fact that the token arrived at the failure state.
\end{proof}

Since these gadgets use exponentially large switching orders, this construction
would have exponential size if written down in the explicit format. Note,
however, that all of the switching orders can be written down in the succinct
format in polynomially many bits. Moreover, the construction has exactly one
switching state for each gate in the circuit, and three extra states for the
control, target, and failure nodes. Every state in the construction can be
created using only the inputs and outputs of the relevant gate in the circuit,
which means that the reduction can be carried out in logarithmic space. Thus, we
have the following:

\begin{thm}
\label{thm:phard}
Deciding the winner of a succinct zero-player RSG is
\P-hard under logspace reductions.
\end{thm}

\subsection{Succinct Zero-Player Games are in UEOPL and CLS}
\label{sec:ueoplzp}

G\"{a}rtner et al.~\cite{GHHKMS18} have shown that the problem of solving an explicit
zero-player game lies in \CLS (which has recently been shown to be equal to
$\PPAD~\cap~\PLS$~\cite{FGHS21}). Their proof reduces the problem to {\sc End-of-Metered-Line}, which is a problem that lies in \CLS~\cite{HY17}.
{\sc End-of-Metered-Line} has also been shown to lie in the recently defined
complexity class \UEOPL~\cite{FGMS20}.

In this section, we show that \emph{succinct} zero-player games also lie in both
\CLS and \UEOPL. We do so by adopting the same strategy as G\"{a}rtner et al.,
namely reducing to {\sc End-of-Metered-Line}.

\paragraph{\bf True and false switching flows.}
The crux of the reduction of G\"{a}rtner et al.\ is a method for differentiating
between \emph{true} and \emph{false} switching flows. A switching flow for a
zero-player game is simply the specialization of a controlled switching flow,
which we defined in \Cref{sec:oneplayer}, in which all nodes are
switching nodes.
This matches the original
definition of a switching flow given by Dohrau et al.~\cite{DGKMW16}.

In this section, it will be convenient to consider switching flows for which the
final node is not necessarily the target of the game. We say that a switching
flow \emph{has target} $x$, if it is a switching flow for a game in which the
target node is $x$.

Since a switching flow is just a specialization of a controlled switching flow,
\Cref{lem:mar2sw} and \Cref{lem:sw2mar} already prove that the reachability
player wins if and only if there is a switching flow, which was already observed
by Dohrau et al.~\cite{DGKMW16}. However, they point out that
not every switching flow corresponds to an actual outcome of the game.

Since we are in the zero-player setting, there is exactly one play of the game.
For each integer~$i$, let $N_i : E \rightarrow \nats \cup \{\infty\}$ be the function that
gives the number of times each edge is used by the first~$i$ steps of 
the play.
We call these functions the \emph{run profiles} of $G$. 

We define $x_i$ to
be the last vertex visited by the first $i$ steps of the play.
It is not difficult to
prove that each $N_i$ is a switching flow with target vertex $x_i$, but the converse is not always true.
Specifically, there can exist switching flows $F$ such that for all $i$ we have
$F(e) \ne N_i(e)$ for at least one edge $e$. 

For example, consider the right-hand diagram in \Cref{fig:csf}. There are
three units of flow leaving~$u$, but if the player follows the marginal strategy
associated with the flow, then the downwards edge from $u$ will only be used
twice. Thus, there is one unit of ``false flow'' between $u$ and the vertex
immediately beneath it. While the diagram depicts a one-player game, 
the phenomenon can also occur in the zero-player setting, as pointed out by
Dohrau et al.~\cite{DGKMW16}.

This leads us to the following
definition.
\begin{defi}[True/False Switching Flows]
A switching flow $F : E \rightarrow \nats$ is said to be true if there exists
and $i$ such that $F(e) = N_i(e)$ for all edges $e \in E$, and it is said to be
false if this is not the case. 
\end{defi}
Note that false switching flows still witness reachability, 
but they do not correctly characterise a run profile of the game.

\paragraph{\bf Detecting false switching flows.}
G\"{a}rtner et al. give the following characterisation of true switching flows.
Their work considered \emph{binary} explicit games, meaning that every vertex
has exactly two outgoing edges.

Given a switching flow $F$ in a binary game, one can easily determine the
\emph{most recently used edge} at each vertex. 
\begin{defi}[Most Recently Used Edge -- Binary Game] 
Suppose that vertex $v$ has two
outgoing edges $(v, u_1)$ and $(v, u_2)$, and that the switching order for $v$
is $\langle u_1, u_2 \rangle$. 
\begin{itemize}
\item if $F(v, u_1) = F(v, u_2) > 0$ then the most recently
used edge is $u_2$, 
\item if $F(v, u_1) = F(v, u_2) + 1$ then the most recently
used edge is $u_1$, and
\item if $F(v, u_1) = F(v, u_2) = 0$ then there is no recently used edge.
\end{itemize}
\end{defi}
\noindent Note that no other possibility is allowed by the definition
of a switching flow.

The MRU graph of a switching flow $F$ in a game $G$ is denoted as
$\mru(G, F)$, and it is obtained by deleting all edges from $G$ that are not a
most recently used edge. Note that every vertex has at most one outgoing edge in
an MRU graph. G\"{a}rtner et al.\ use the MRU graph to give the following
characterisation of false switching flows.

\begin{lem}[G{\"a}rtner et al.~\cite{GHHKMS18}]
\label{lem:falsesf}
Let $F$ be a switching flow for a binary explicit zero-player switching game
with target~$x$.
Then $F$ is a true switching flow 
if and only if one of the following conditions holds.
\begin{enumerate}
\item $\mru(G, F)$ is acyclic.
\item There is exactly one cycle in $\mru(G, F)$, and the target node $x$ lies on this cycle.
\end{enumerate}
\end{lem}

\noindent The idea here is that, if the MRU graph of $F$ contains a cycle that
does not contain the target node, then that cycle contains false flow.
Specifically, for
each edge $e$ on that cycle, there will be no index $i$ such that $F(e) =
N_i(e)$. Conversely, the non-existence of such a cycle is sufficient to ensure
that $F$ is a true switching flow.
We will extend this lemma to cover succinct non-binary zero-player games.


\paragraph{\bf Generalising most-recently used edges.}
The definition of a most-recently used edge can be generalised to non-binary
succinct games in a natural way.
\begin{defi}[Most Recently Used Edge -- Succinct Non-Binary-Game]
Let $F$ be a switching flow, let $v$ be a vertex, and let 
$\langle (u_1, t_1), (u_2, t_2), \dots, (u_m, t_m) \rangle$ be the switching
order at $v$, which may be succinct or explicit.  If $t = \sum_{(w, v)\in E}
F(w, v)$ denotes the total amount of flow incoming at~$v$, and $k = \sum_{i=1}^m
t_i$ is the length of the switching order at $v$, then we use the following
definitions.
\begin{itemize}
\item If $t = 0$, then there is no most recently used edge at $v$.
\item If $t > 0$ and $t \bmod k = 0$, then the most recently used edge
at $v$ is $u_m$.
\item If $t > 0$ and $t \bmod k = i$, then the most recently used edge
at $v$ is the vertex $u$ that appears at position $i-1$ in $\ord(v)$.
\end{itemize}
\end{defi}
\noindent Observe that, even for succinctly represented orderings, we can
compute the most recently used edge at each vertex in polynomial time.

Recall that, given a flow $F$ in a game $G$, the graph $\mru(G, F)$ contains the
set of most recently used edges in $F$. This definition also applies to
non-binary succinct games, using the definition of a most-recently used edge
given above.

\paragraph{\bf Non-binary explicit games.}

We begin by slightly generalising \Cref{lem:falsesf} to non-binary explicit
games. In fact, the proof of G\"{a}rtner et al.\ essentially already works for
non-binary games, but several details need to be updated, and so we adapt the proof
here ourselves for the sake of completeness.

\begin{lem}
\label{lem:falsesf2}
Let $F$ be a switching flow for an explicit zero-player switching game with
target node~$x$.
Then $F$ is a true switching flow if and only if one of the following conditions holds.
\begin{enumerate}
\item $\mru(G, F)$ is acyclic.
\item There is exactly one cycle in $\mru(G, F)$, and the target node $x$ lies on this cycle.
\end{enumerate}
\end{lem}
\begin{proof}
We prove the directions separately.
\begin{description}
\item[The $\Rightarrow$ direction]
For the forward direction, we must show that if $F$ is a true switching flow,
then the MRU graph of $F$ satisfies the conditions. Let $i$ be the index such
that $F = N_i$. Observe that the MRU graph of $F$ is the same as the MRU graph
of~$N_i$. We will show that the conditions hold for $N_i$ by induction on $i$.


For the base case, observe that $N_0$ corresponds to the prefix of the run that
has not used any edges, and so the MRU graph of $N_0$ is empty, and thus
acyclic. For the inductive step, suppose that the MRU graph of $N_j$ satisfies
the
conditions, and that the $j+1$th step of the play moves from $v$ to $u$. This
transforms the MRU graph in the following way: the existing edge at $v$ is
deleted, and the new most-recently used edge of $v$ is set to $(v, u)$. If the
MRU graph of $N_j$ contains a cycle, then by the inductive hypothesis it must
pass through $v$. Therefore, deleting the outgoing edge of $v$ makes the graph
acyclic. Adding the edge $(v, u)$ to the graph may introduce a new cycle, but
this passes through $u$, which is the target vertex of~$N_{j+1}$. Hence, the MRU
graph of~$N_{j+1}$ satisfies the conditions.

To conclude the forward direction of the proof, we have shown that each run
profile~$N_{i}$ satisfies the conditions of the lemma. Since $F = N_{i}$ for
some $i$, this means that $F$ also satisfies the conditions of the lemma.



\item [The $\Leftarrow$ direction]
For the other direction, we must show that if the conditions on the MRU graph
are satisfied, then $F$ is a true switching flow. For the sake of contradiction,
we suppose that this is not the case. Let $i$ be the largest index such that
$N_i(e) \le F(e)$ for all edges $e$, and let~$\delta$ be defined so
that $\delta(e) = F(e) - N_i(e)$ for all edges $e$. 

Let $x$ be the target node of $N_i(e)$. The first observation is that $\delta(x,
u) = 0$ for all edges $(x, u)$.  To see why, observe that if $\sum_{(v, x) \in
E} N_i(v, x) < \sum_{(v, x) \in E} F(v, x)$, ie., if $x$ receives less total
flow under $N_i$ than it does under $F$, then the switching flow $N_{i+1}$ will
also satisfy $N_i(e) \le F(e)$ for all edges $e$. On the other hand, if
$\sum_{(v, x) \in E} N_i(v, x) = \sum_{(v, x) \in E} F(v, x)$, then the
switching flow constraints force both flows to send the same amount of flow to each
outgoing edge of $x$, which implies that $\delta(x, u) = 0$ for all edges $(x,
u)$.

If $\delta(e) = 0$ for all edges $e$, then we have a contradiction with the
assumption that~$F$ is a false switching flow. Otherwise, we will prove that
there is a cycle in the MRU graph of $F$ that only uses edges $e$ with
$\delta(e) > 0$. 

First note that if $\delta(v, u) > 0$ for some edge $(v, u)$,
then there must be an edge $(u, w)$ satisfying $\delta(u, w) > 0$. This is
because the flow constraints ensure that $\sum_{(x, u) \in E} F(x, u) = \sum_{(u, y) \in E} F(u, y)$ 
and also $\sum_{(x, u) \in E} N_i(x, u) = \sum_{(u, y) \in E} N_i(u, y)$. Since
we have by assumption that $F(v, u) > N_i(v, u)$, and since $F(e) \ge N_i(e)$
for all edges $e$, there must be an outgoing edge at $u$ satisfying $F(u, w) >
N_i(u, w)$. 


Next we show that if $v$ has an outgoing edge $e$ satisfying $\delta(e)
> 0$, then if $e'$ denotes the most-recently used edge at $v$, then we also have
 that $\delta(e') > 0$. 
Suppose that the switching order at $v$ is $\langle u_1, u_2,
\dots, u_k \rangle$, and suppose that $\delta(v, u_j) > 0$ for some $j$. There
are two cases to consider.
\begin{itemize}
\item If $F(v, u_1) = F(v, u_2) = \dots = F(v, u_k)$, that is, if $F$ sends the
same amount of flow to all successors of $v$, then we must
have $F(v, u_k) > N_i(v, u_k)$ due to the fact that $\delta(v, u_j) > 0$, and
the fact that the switching flow constraints ensure that flow is sent to $u_j$ before it is
sent to $u_k$. Since the
MRU graph uses $u_k$ at $v$ by definition, this implies that 
MRU graph uses an edge at $v$ with $\delta(e) > 0$.

\item If $F(v, u_1) = \dots = F(v, u_\ell) = F(v, u_{\ell+1}) + 1 = \dots = F(v, u_k)$, 
that is, if the flow sends one unit of flow more to vertices $u_1$ through
$u_\ell$, 
then observe the following. 
\begin{itemize}
\item If $j < \ell$ then as before, since $\delta(v, u_j) > 0$, and the fact that
the flow constraints send flow to $u_j$ before $u_\ell$, we have $\delta(v, u_\ell) >
0$.
\item If $j = \ell$ then $\delta(v, u_\ell) = \delta(v, u_j) > 0$ by assumption.
\item If $j > \ell$ then since $N_i(u_j) \le N_i(u_\ell)$, we have that 
\begin{equation*}
F(v, u_\ell) - N_i(v, u_\ell) \ge F(v, u_j) - N_i(v, u_j) > 0.
\end{equation*}
\end{itemize}
The MRU graph uses $u_\ell$ at $v$ by definition, and so we again have shown that
the MRU graph uses an edge at $v$ with $\delta(e) > 0$.
\end{itemize}

We can now prove that the MRU graph of $F$ contains a cycle that
that only uses edges $e$ with
$\delta(e) > 0$. We can construct this cycle in the following way.

\begin{enumerate}
\item Start with an arbitrary edge $(v, u)$ satisfying $\delta(v, u) > 0$. 
\item Identify the most recently used edge $(v, w)$ at $v$. We have proved that
$\delta(v, w) > 0$. 
\item We have shown that since $\delta(v, w) > 0$, there exists an outgoing edge
$(w, x)$ satisfying $\delta(w, x) > 0$. 
\item Repeat the argument starting from Step 2 with the edge $(w, x)$. 
\end{enumerate}
This argument constructs a path of most-recently used edges such that each edge
$e$ satisfies $\delta(e) > 0$. Note that the path can never visit the target
vertex, since we have shown that all outgoing edges $e$ of the target satisfy
$\delta(e) = 0$. Since the graph is finite, this path must
eventually loop back to itself, which creates a cycle. Thus, the MRU graph
contains a cycle that does not contain the target vertex, which gives a
contradiction. \qedhere
\end{description}
\end{proof}

\paragraph{\bf Succinct games.}
We can now prove that the same property holds for succinct games. Note that none
of the properties in \Cref{lem:falsesf2} care about the size of the game. So, for each vertex $v$ in a succinct game $G$, we can take a succinct ordering 
$\ord(v) = \langle (u_1, t_1),
(u_2, t_2), \dots, (u_k, t_k) \rangle$, and blow it up to the explicit switching order
\begin{equation*}
\rep(u_1, t_1) \cdot \rep(u_2, t_2) \cdot \; \dots \; \cdot \rep(u_k, t_k)
\end{equation*}
to obtain a (potentially exponentially large) explicit game $G'$. We can then
apply 
\Cref{lem:falsesf2} to $G'$, and observe that an edge is most recently
used in $G'$ if and only if it is most recently used in $G$, since we have not
actually changed the switching order at any vertex. So, we obtain the following:
\begin{lem}
\label{lem:fsfsuccinct}
Let $F$ be a switching flow for a succinct zero-player switching game with
target node~$x$.
Then $F$ is a true switching flow if and only if one of the following conditions holds.
\begin{enumerate}
\item $\mru(G,F)$ is acyclic.
\item There is exactly one cycle in $\mru(G,F)$, and the target node $x$ lies on this
cycle.
\end{enumerate}
\end{lem}

\paragraph{\bf UEOPL and CLS containment.}
G\"{a}rtner et al.\ rely on the properties of \Cref{lem:falsesf} to prove
their containment result~\cite{GHHKMS18}. Having shown the analogue of that
lemma for succinct games, we can now follow their proof directly. 
We summarise
the technique here, and direct the reader to G\"{a}rtner et al.~\cite{GHHKMS18} for the full
details.

Since the game is zero-player, there is a unique play $\pi = (v_1, C_1), (v_2,
C_2), \dots, (v_k, C_k)$ with $v_1 = s$ and $v_k = t$. For each $i$, the run
profile $N_i$ gives is the unique true switching flow that witnesses that the
play passes through $(v_i, C_i)$ after $i$ steps.
Hence, we can build an (exponentially long) line of switching flows $N_0, N_1,
\dots, N_k$. Given a switching flow $N_i$, we can compute the next switching
flow $N_{i+1}$, and the previous switching flow $N_{i-1}$. This is enough to
build an {\sc End-of-Metered-Line} instance, which gives the following result.


\begin{thm}
\label{thm:ueopl}
The problem of solving a zero-player succinct switching game is in \UEOPL and
\CLS.
\end{thm}

\section{Further work}
\label{sec:fw}

Many interesting open problems remain. For the zero-player case in the \emph{explicit}
case, there is an
extremely large gap between the upper bounds of \NPcapcoNP and \PLS and the easy
lower bound of \NL that we showed here.  We conjecture that the problem is in
fact \P-complete, but despite much effort, we were unable to improve upon the
upper or lower bounds.

For the one-player case we have shown tight bounds. For the two-player case we
have shown a lower bound of \PSPACE and an upper bounds of \EXPTIME. We
conjecture that the lower bound can be strengthened, since we did not make
strong use of the memory requirements that we identified in
\Cref{sec:memory}.

Finally, here we studied the problem of reachability, which is one of the
simplest model checking tasks. What is the complexity of model checking more
complex specifications?

\section*{Acknowledgements}
Fearnley was supported by EPSRC grant EP/P020909/1 ``Solving Parity Games in
Theory and Practice''.
Mnich was supported by ERC Starting Grant 306465 (BeyondWorstCase) and DFG grant MN 59/4-1.
A Visiting Fellowship from the Department of Computer Science
at the University of Liverpool allowed Mnich to visit Fearnley, Gairing, and Savani.

\bibliographystyle{alpha}
\bibliography{references}

\newcommand{\etalchar}[1]{$^{#1}$}
\begin{thebibliography}{DGK{\etalchar{+}}17}

\bibitem[AB13]{AB13}
Hoda Akbari and Petra Berenbrink.
\newblock Parallel rotor walks on finite graphs and applications in discrete
  load balancing.
\newblock In {\em Proc.\ of SPAA 2013}, pages 186--195, 2013.

\bibitem[CDFS10]{CDFS10}
Joshua Cooper, Benjamin Doerr, Tobias Friedrich, and Joel Spencer.
\newblock Deterministic random walks on regular trees.
\newblock {\em Random Structures Algorithms}, 37(3):353--366, 2010.

\bibitem[CDST07]{CDST07}
Joshua Cooper, Benjamin Doerr, Joel Spencer, and G\'abor Tardos.
\newblock Deterministic random walks on the integers.
\newblock {\em European J. Combin.}, 28(8):2072--2090, 2007.

\bibitem[CKS81]{CKS81}
Ashok~K. Chandra, Dexter~C. Kozen, and Larry~J. Stockmeyer.
\newblock Alternation.
\newblock {\em J. Assoc. Comput. Mach.}, 28(1):114--133, 1981.

\bibitem[Con89]{Con89}
Anne Condon.
\newblock {\em Computational models of games}.
\newblock {ACM} distinguished dissertations. {MIT} Press, 1989.

\bibitem[Con92]{Con92}
Anne Condon.
\newblock The complexity of stochastic games.
\newblock {\em Inf. Comput.}, 96(2):203--224, 1992.

\bibitem[CS06]{CS06}
Joshua Cooper and Joel Spencer.
\newblock Simulating a random walk with constant error.
\newblock {\em Combin. Probab. Comput.}, 15(6):815--822, 2006.

\bibitem[DF09]{DF09}
Benjamin Doerr and Tobias Friedrich.
\newblock Deterministic random walks on the two-dimensional grid.
\newblock {\em Combin. Probab. Comput.}, 18(1-2):123--144, 2009.

\bibitem[DGK{\etalchar{+}}16]{DGKMW16}
J{\'{e}}r{\^{o}}me Dohrau, Bernd G{\"{a}}rtner, Manuel Kohler, Ji{\v r}{\'\i}
  Matou\v{s}ek, and Emo Welzl.
\newblock {ARRIVAL}: A zero-player graph game in ${\NP} \cap {\coNP}$.
\newblock Technical report, 2016.
\newblock \url{https://arxiv.org/abs/1605.03546}.

\bibitem[DGK{\etalchar{+}}17]{DGKMW17}
J{\'e}r{\^o}me Dohrau, Bernd G{\"a}rtner, Manuel Kohler, Ji{\v r}{\'\i}
  Matou{\v s}ek, and Emo Welzl.
\newblock A{RRIVAL}: a zero-player graph game in {$\NP\cap \coNP$}.
\newblock In {\em A journey through discrete mathematics}, pages 367--374.
  Springer, Cham, 2017.

\bibitem[FGHS21]{FGHS21}
John Fearnley, Paul Goldberg, Alexandros Hollender, and Rahul Savani.
\newblock The complexity of gradient descent: {\CLS} = {\PPAD}\,$\cap$\,{\PLS}.
\newblock In {\em Proc.\ of {STOC} 2021}, 2021.

\bibitem[FGMS18]{FGMS18b}
John Fearnley, Martin Gairing, Matthias Mnich, and Rahul Savani.
\newblock Reachability switching games.
\newblock In {\em Proc.\ of ICALP 2018}, volume 107 of {\em Leibniz Int. Proc.
  Informatics}, pages 124:1--124:14, 2018.

\bibitem[FGMS20]{FGMS20}
John Fearnley, Spencer Gordon, Ruta Mehta, and Rahul Savani.
\newblock Unique end of potential line.
\newblock {\em J. Comput. Syst. Sci.}, 114:1--35, 2020.

\bibitem[FGS12]{FGS12}
Tobias Friedrich, Martin Gairing, and Thomas Sauerwald.
\newblock Quasirandom load balancing.
\newblock {\em SIAM J. Comput.}, 41(4):747--771, 2012.

\bibitem[GHH{\etalchar{+}}18]{GHHKMS18}
Bernd G\"{a}rtner, Thomas~Dueholm Hansen, Pavel Hub\'{a}\v{c}ek, Karel
  Kr\'{a}l, Hagar Mosaad, and Veronika Sl\'{i}vov\'{a}.
\newblock {ARRIVAL}: Next stop in {CLS}.
\newblock In {\em Proc.\ of ICALP 2018}, volume 107 of {\em Leibniz Int. Proc.
  Informatics}, pages 60:1--60:13, 2018.

\bibitem[GHH21]{GHH21}
Bernd G{\"{a}}rtner, Sebastian Haslebacher, and Hung~P. Hoang.
\newblock A subexponential algorithm for {ARRIVAL}.
\newblock Technical report, 2021.
\newblock \url{https://arxiv.org/abs/2102.06427}.

\bibitem[GHR95]{GHR95}
Raymond Greenlaw, H.~James Hoover, and Walter~L. Ruzzo.
\newblock {\em Limits to parallel computation: {P}-completeness theory}.
\newblock The Clarendon Press, Oxford University Press, New York, 1995.

\bibitem[GP09]{GP09}
Jan~Friso Groote and Bas Ploeger.
\newblock Switching graphs.
\newblock {\em Internat. J. Found. Comput. Sci.}, 20(5):869--886, 2009.

\bibitem[HLM{\etalchar{+}}08]{HLMPP08}
Alexander~E. Holroyd, Lionel Levine, Karola M\'esz\'aros, Yuval Peres, James
  Propp, and David~B. Wilson.
\newblock Chip-firing and rotor-routing on directed graphs.
\newblock In {\em In and out of equilibrium. 2}, volume~60 of {\em Progr.
  Probab.}, pages 331--364. Birkh\"auser, Basel, 2008.

\bibitem[HP10]{HP10}
Alexander~E. Holroyd and James Propp.
\newblock Rotor walks and {M}arkov chains.
\newblock In {\em Algorithmic probability and combinatorics}, volume 520 of
  {\em Contemp. Math.}, pages 105--126. Amer. Math. Soc., Providence, RI, 2010.

\bibitem[HY17]{HY17}
Pavel Hub\'{a}\v{c}ek and Eylon Yogev.
\newblock Hardness of continuous local search: Query complexity and
  cryptographic lower bounds.
\newblock In {\em Proc.\ of SODA 2017}, pages 1352--1371, 2017.

\bibitem[{Kar}17]{Kar17}
{Karthik {C. S.}}
\newblock Did the train reach its destination: The complexity of finding a
  witness.
\newblock {\em Inf. Process. Lett.}, 121:17--21, 2017.

\bibitem[KNP11]{KNP11}
Marta Kwiatkowska, Gethin Norman, and David Parker.
\newblock {PRISM} 4.0: Verification of probabilistic real-time systems.
\newblock In {\em Proc.\ of CAV 2011}, volume 6806 of {\em Lecture Notes
  Comput. Sci.}, pages 585--591, 2011.

\bibitem[KRW12]{KRW12}
Bastian Katz, Ignaz Rutter, and Gerhard Woeginger.
\newblock An algorithmic study of switch graphs.
\newblock {\em Acta Inform.}, 49(5):295--312, 2012.

\bibitem[Mei89]{Mei89}
Christoph Meinel.
\newblock Switching graphs and their complexity.
\newblock In {\em Proc.\ of MFCS 1989}, volume 379 of {\em Lecture Notes
  Comput. Sci.}, pages 350--359, 1989.

\bibitem[PDDK96]{PDDK96}
Vyatcheslav~B. Priezzhev, Deepak Dhar, Abhishek Dhar, and Supriya
  Krishnamurthy.
\newblock Eulerian walkers as a model of self-organized criticality.
\newblock {\em Phys. Rev. Lett.}, 77(25):5079, 1996.

\bibitem[PT87]{PT87}
Christos~H. Papadimitriou and John~N. Tsitsiklis.
\newblock The complexity of markov decision processes.
\newblock {\em Math. Oper. Res.}, 12(3):441--450, 1987.

\bibitem[Put94]{Put94}
Martin~L. Puterman.
\newblock {\em Markov Decision Processes: Discrete Stochastic Dynamic
  Programming}.
\newblock John Wiley \& Sons, Inc., New York, NY, USA, 1st edition, 1994.

\bibitem[Rei09]{Rei09}
Klaus Reinhardt.
\newblock The simple reachability problem in switch graphs.
\newblock In {\em Proc.\ of SOFSEM 2009}, pages 461--472, 2009.

\bibitem[Tov84]{Tovey1984}
Craig~A. Tovey.
\newblock A simplified \texttt{NP}-complete satisfiability problem.
\newblock {\em Discrete Appl. Math.}, 8(1):85--89, 1984.

\end{thebibliography}

\end{document}